\documentclass[usenatbib]{mn2e}
\usepackage{epsfig,journals}

\newcommand	{\cm}{\ifmmode {\rm cm}\else cm\fi}
\newcommand	{\km}{\ifmmode {\rm km}\else km\fi}
\newcommand	{\pc}{\ifmmode {\rm pc}\else pc\fi}
\newcommand	{\kpc}{\ifmmode {\rm kpc}\else kpc\fi}
\newcommand	{\Mpc}{\ifmmode {\rm Mpc}\else Mpc\fi}
\newcommand	{\ly}{\ifmmode {\rm ly}\else ly\fi}
\newcommand	{\s}{\ifmmode {\rm s}\else s\fi}
\newcommand	{\yr}{\ifmmode {\rm yr}\else y\fi}
\newcommand	{\K}{\ifmmode {\rm K}\else K\fi}
\newcommand	{\erg}{\ifmmode {\rm erg}\else erg\fi}
\newcommand	{\dyn}{\ifmmode {\rm dyn}\else dyn\fi}
\newcommand	{\Msun}{\ifmmode {M}_{\mathord\odot}\else 
			$M_{\mathord\odot}$\fi}
\newcommand	{\msun}{\ifmmode {M}_{\mathord\odot}\else 
			$M_{\mathord\odot}$\fi}
\newcommand{\rhoa}{\rho_{a}}
\newcommand{\tff}{t_{\rm ff}}
\newcommand{\Ein}{{E_{\rm in}}}

\newcommand	{\Rsun}{\ifmmode {R_\odot}\else R$_\odot$ \fi}
\newcommand	{\AU}{\ifmmode {{\rm AU}}\else AU \fi} 
\newcommand	{\kms}{\ifmmode{\rm km\;s^{-1}}\else km~s\e \fi}
\newcommand	{\kps}{\ifmmode{\rm km\;s^{-1}}\else km~s\e \fi}

\newcommand{\nat}{Nature}
\newcommand{\physrep}{Physics Reports}

\newcommand{\Eisogam}{{E}_{\gamma, \rm iso}}
\newcommand{\Egam}{{E}_{\gamma}}
\newcommand{\Etot}{{E}}
\newcommand{\Efpf}{E_{\gamma,50.5}}
\newcommand{\Lt}{{\tilde{L}}}

\newcommand{\Liso}{L_{\rm iso}}
\newcommand{\Lisogam}{L_{\gamma \rm iso}}
\newcommand{\Lisofto}{L_{\gamma \rm iso,51}}
\newcommand{\Lgam}{L_\gamma}
\newcommand{\Ltot}{L}
\newcommand{\tobs}{t_{\rm obs}}
\newcommand{\tgam}{t_{\gamma, \rm obs}}
\newcommand{\rIS}{r_{\rm IS}}
\newcommand{\g}{{\rm g}}
\newcommand{\Ma}{M}
\newcommand{\Menv}{M_{\rm env}}

\newcommand{\lesssim}{{\la}}
\newcommand{\gtrsim}{{\ga}}
\begin{document}

\title[SN Hosts for GRB Jets: Dynamical Constraints]
{Supernova Hosts for Gamma-Ray Burst Jets: Dynamical
Constraints} \author[C. D. Matzner]{Christopher D. Matzner$^{1,2}$ \\
	$^1$Canadian Institute for Theoretical Astrophysics, 
     University of Toronto \\
     $^2$ Present address: Dept. of Astronomy and Astrophysics,
     University of Toronto, 60 St.~George Street, 
     Toronto, ON M5S 3H8, Canada} 
\date{Received xxxx, 2002}
\pagerange{\pageref{firstpage}--\pageref{lastpage}}
\pubyear{2002}\maketitle\label{firstpage}

\begin{abstract}
I constrain a possible supernova origin for gamma-ray bursts by
modeling the dynamical interaction between a relativistic jet and a
stellar envelope surrounding it.  The delay in observer's time
introduced by the jet traversing the envelope should not be long
compared to the duration of gamma-ray emission; also, the jet should
not be swallowed by a spherical explosion it powers.  The only stellar
progenitors that comfortably satisfy these constraints, if one assumes
that jets move ballistically within their host stars, are compact
carbon-oxygen or helium post-Wolf-Rayet stars (type Ic or Ib
supernovae); type II supernovae are ruled out.  Notably, very massive
stars do not appear capable of producing the observed bursts at any
redshift unless the stellar envelope is stripped prior to collapse.
The presence of a dense stellar wind places an upper limit on the
Lorentz factor of the jet in the internal shock model; however, this
constraint may be evaded if the wind is swept forward by a photon
precursor.  Shock breakout and cocoon blowout are considered
individually; neither presents a likely source of precursors for 
cosmological GRBs.   

These envelope constraints could conceivably be circumvented if jets
are laterally pressure-confined while traversing the outer stellar
envelope.  If so, jets responsible for observed GRBs must either have
been launched from a region several hundred kilometers wide, or have mixed
with envelope material as they travel.  A phase of pressure confinement
and mixing would imprint correlations among jets that may explain
observed GRB variability-luminosity and lag-luminosity correlations.
\end{abstract}


\section{Introduction}\label{Intro}
There is presently a growing body of circumstantial
evidence linking some long-duration gamma ray bursts (GRBs) with
afterglows to the explosions of massive stars.  Supernovae (SN) or
supernova-like features have been identified in seven afterglows
\citep{1998Natur.395..670G, 1999Natur.401..453B, 1999ApJ...521L.111R,
2000ApJ...534L..57T, 2001ApJ...552L.121B, 2001A&A...378..996L}
although some of these could be light echoes from dust clouds
\citep{2000ApJ...534L.151E}.
The afterglows of six other GRBs have been interpreted in terms of a
wind-like ambient medium as expected around a massive star at the end
of its life \citep{1999ApJ...525L..81F,2001A&A...371...52H,
2000ApJ...534..559F, 2001ApJ...551..940L,2002ApJ...572L..51P}.
However these are often equally well explained by a collimated flow in
a uniform medium.  X-ray lines have been detected with moderate
confidence in about half the afterglows for which they were
investigated \citep{ToshioDaisuke2000}; however, the data analysis has
been questioned \citep[see][]{2003MNRAS.339..600R}.  If real, these
are most easily explained by dense material surrounding the burst
engine, suggesting a stellar origin
\citep[e.g.,][]{2001RMxAC..10..182B}.

\cite{2001ApJ...562L..55F} have recently derived beaming angles for a
number of GRBs from observations of their afterglows.  These authors
derive gamma-ray energies reminiscent of supernovae: roughly $3\times
10^{50}$ erg for the observed lobes of GRBs.  Similar results were
reported by \cite{2001ApJ...554..667P} and \cite{2001ApJ...547..922F}.

Clear evidence that GRBs occur very close to massive star formation
would be almost as conclusive as a SN signature in an individual GRB.
Several GRB afterglows show evidence for high column densities (980703
and 980329; \citealt{2001ApJ...549L.209G}) or high local gas densities
(000926 and 980519; \citealt{2001ApJ...559..123H},
\citealt{2000MNRAS.317..170W}), both of which connote star-forming
regions.
Likewise, the intrinsic extinction of GRB 000926 is characteristic of
a galaxy disk \citep{2001ApJ...549L...7P}.  \cite{2002AJ....123.1111B}
have shown that the observed locations within hosts imply a tight
correlation between GRBs and stellar populations, considered too tight
\citep{1999MNRAS.305..763B} to be matched by merging neutron stars.
Note however that the locations of merging neutron star pairs depends
on their uncertain distribution of natal kicks.

If GRBs are a rare byproduct of star formation, rapidly star-forming
galaxies should be over-represented as GRB hosts.  In optical light
host galaxies tend to look ordinary compared to contemporaries in the
Hubble Deep Field, but [Ne III] and [O II] and infrared observations
often indicate elevated star formation rates
\citep{2001grba.conf..218D}.  At least eight afterglows have been
associated with starburst or interacting galaxies
\citep{2001ApJ...562..654D,2002ApJ...565..829F,2002ApJ...566..229C}.

Although the association between long-duration GRBs and SNe is
tentative (and applies only to the long-duration bursts for which 
afterglows are observed), the above evidence warrants a careful
evaluation.  There are two ways a SN can create a
GRB.  \cite{1968CaJPh..46..476C} predicted that gamma rays might be
produced in the very fastest, outermost ejecta of an ordinary
supernova explosion.  This proposal was recently revived by
\cite{2000ApJ...545..364M} and \cite{2001ApJ...551..946T}.  These
authors showed that the GRB (980425) most compellingly associated with
a SN (1998bw) is likely to be the result of trans-relativistic SN
ejecta colliding with a stellar wind \citep[see
also][]{2001gra..conf..200M}.  In their model, as conjectured by
\cite{1998Natur.395..672I}, SN 1998bw was spherically symmetric or
mildly asymmetric, and produced the GRB in an external shock.
Scaled-up versions of this model could produce external shock GRBs, at
the expense of vast amounts ($\sim 10^{54}$ erg) of energy in
nonrelativistic ejecta.  

In contrast, \cite{1995ApJ...455L.143S} have argued that most GRBs
require internal emission (e.g., by internal shocks) within unsteady
ultrarelativistic winds or jets, as originally suggested by
\cite{1994ApJ...430L..93R}.  The arguments for internal emission are
strongest for rapidly-fluctuating cosmological bursts with hard
spectra, those least resembling GRB 980425; also, see
\cite{1999ApJ...513L...5D} for arguments in support of external
shocks.  I shall assume for the purposes of this investigation that
cosmological GRBs involve internal emission within optically thin,
ultrarelativistic outflows.  For this to result from a SN, a jet
must emanate from a star's core and pierce its envelope -- shedding
the baryons in its path -- prior to producing the gamma rays observed
at Earth.  Such a jetlike explosion is the conventional model
\citep[e.g.,][]{1999ApJ...524..262M} for a supernova origin of
cosmological GRBs. 

The goal of this paper will be to develop analytical models for the
phase of this latter model in which a jet, already created by the
stellar core, must traverse the envelope and shove aside material in
its path.  These models, which are complementary to numerical
simulations \citep{2000ApJ...531L.119A}, are meant to elucidate under
what conditions the hypothesis of a stellar origin is viable for the
observed GRBs.  In \S\S \ref{Inside Star} and
\ref{EnvelopeConstraints} I assume that jets travel ballistically
within their stars; this allows one to place strict constraints on
stellar envelopes.  In \S \ref{Widening} this assumption is
reconsidered.  It is shown that a phase in which the jet is hot, 
pressure confined, and mixing with its environs would have interesting
consequences. 

\subsection{Stellar Progenitors}\label{S:Progenitors}

\begin{figure*}
\centerline{\epsfig{figure=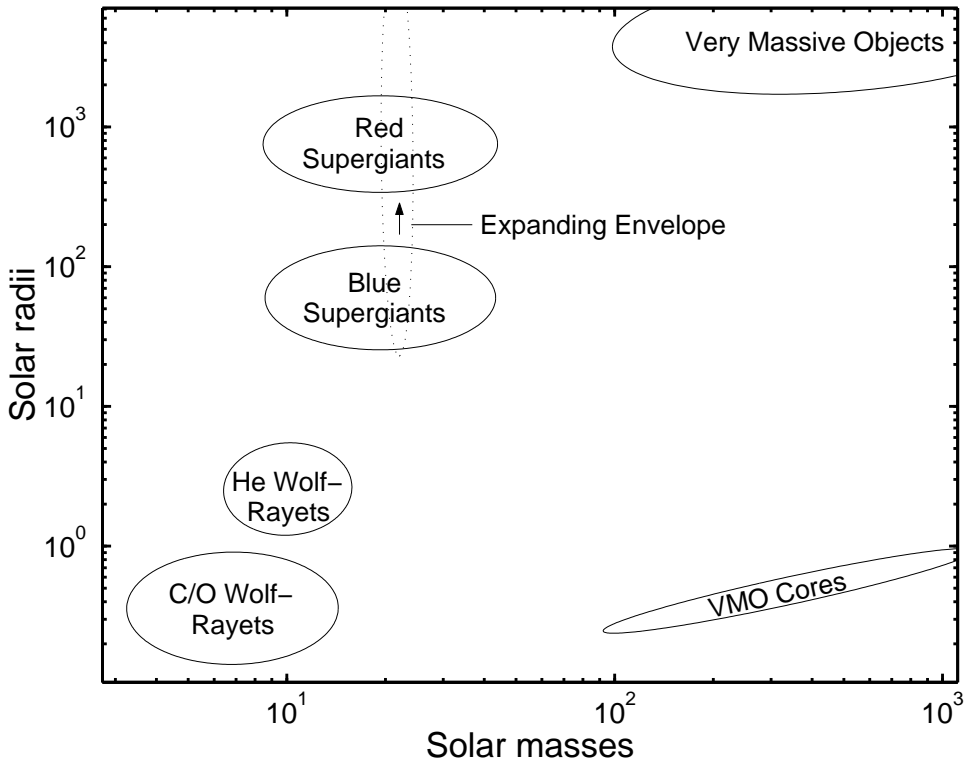,height=4in}} 
\caption[Progenitor masses and radii] {\footnotesize Schematic of the
typical masses and radii of the stellar progenitors considered in this
paper (\S \ref{S:Progenitors}).  The plotted regions each represent several
pre-supernova stellar models from the literature. \label{fig1} }
\end{figure*}

Figure \ref{fig1} sketches the typical masses and radii of the
stellar GRB progenitor candidates considered in this paper. In general,
those that retain an outer envelope (e.g., supergiants) have quite
large radii ($R>10~\Rsun$) at the time of core collapse, whereas those
depleted in hydrogen due to winds or binary interaction (e.g., those
that have been through a Wolf-Rayet phase) are quite compact
($R<10~\Rsun$). Among post-Wolf-Rayet stars, those containing 
helium (``He Wolf-Rayets'' on the plot) are less compact than their
He-depleted peers (``C/O Wolf-Rayets''). 

Very massive objects (VMOs) might have formed at high redshift due to
the difficulty of cooling in the absence of metals and thus the 
large Jeans mass in primordial gas 
\citep[][]{1977Ap&SS..48..145H}.  VMOs may also form today in rare conditions.
Those initially more massive than $\sim 250 \Msun$ die when their
cores collapse to black holes, and are candidates for producing GRBs;
\cite{1984ApJ...280..825B} discuss their evolution.  Again, the pre-collapse
radii of these stars depend on their mass loss. If present, their H
envelope is quite diffuse \citep[see][]{2001ApJ...550..372F}.
Otherwise the remaining convective core is very compact at the point
of collapse.  In addition to winds and binary mass transfer,
VMOs may shed their envelopes in a super-Eddington phase during helium
core contraction. Lone VMOs are thought to retain H envelopes if they
are formed from sufficiently low-metallicity gas \citep[although this
is uncertain;][]{1984ApJ...280..825B}.

The ``supranova'' model of \cite{1999ApJ...527L..43V} posits that a SN
explosion produces a rapidly spinning neutron star massive enough to
collapse after shedding its angular momentum \citep{1999ApJ...526..941B} along
with $\sim 10^{53}$ erg of rotational energy.  The object collapses to
a black hole with an accretion torus, firing a jet through the
pulsar's wind nebula and its sheath of stellar ejecta.
Because of the short viscous times of such tori, this model is most
appropriate for short GRBs rather than the long bursts for which there
is evidence of a SN connection.  Also, we will see in \S
\ref{EnvelopeConstraints} that the expanding stellar ejecta would
prevent the production of a GRB entirely, unless it has either become
Thompson optically thin or has been cleared aside by the breakout of
the pulsar wind nebula.  The latter is possible in this model, because
the pulsar spindown energy exceeds the typical kinetic energy of SN
ejecta.  

\cite{1998ApJ...502L...9F} and \cite{2001ApJ...550..357Z} discuss a
scenario in which the stellar core collapse leading to a GRB results
from the coalescence of a helium Wolf-Rayet star with a compact
companion.  Similarly, \cite{1999ApJ...520..650F} argue that binary
mergers are likely to dominate the production stars whose cores
collapse to black holes accreting through a disk.  Progenitors created
in this fashion will typically be stripped of their outer envelopes,
hence compact.  However the stripped envelope poses a potential
barrier to jet propagation, as in the supranova model, unless it is
strictly confined to the equator of the system or has expanded to the
point of being optically thin. 

In sections \ref{S:freefalltimes} and \ref{S:PressureConf} I examine the
properties of stellar cores at the point of collapse. For this I shall
assume that collapse sets in at the oxygen ignition temperature, $T_9
\equiv T/10^9~\K \simeq 3.2$ \citep[e.g.,][]{1984ApJ...280..825B}. 

\section{Jets Within Stars and Stellar Winds}\label{Inside Star}
\begin{figure*}
\centerline{\epsfig{figure=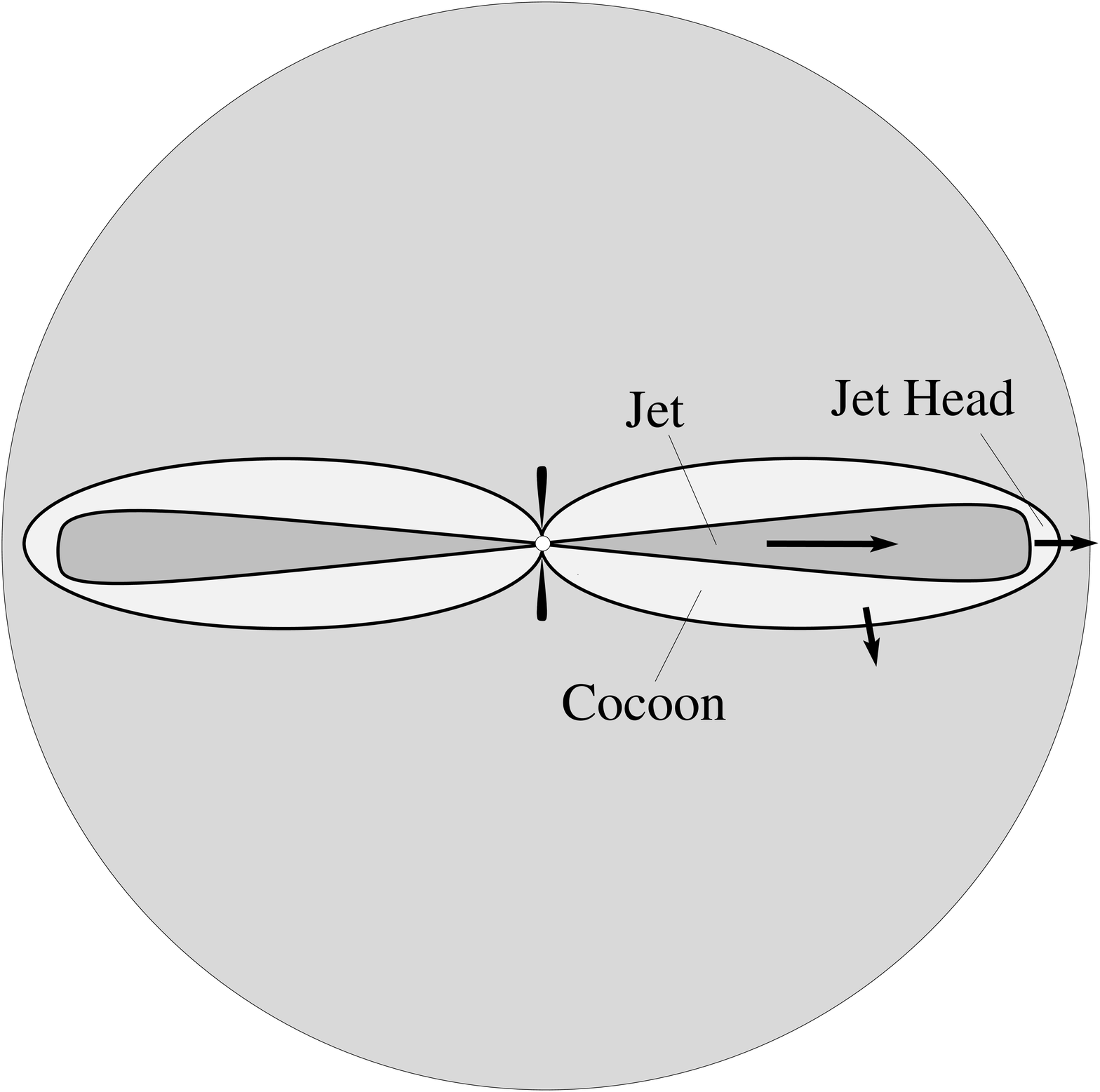, height=4in}} 
\caption[Jet-cocoon structure] {\footnotesize Schematic of a jet
traversing a stellar envelope. The two collide at the jet head, from
which material flows sideways into the cocoon. \label{fig:diagram} }
\end{figure*}

Consider the progress of a relativistic jet outward from a star's core
through the stellar envelope.  Schematically, three distinct regions
develop: the propagating jet; the {\em head} of the jet, where jet
material impacts the stellar envelope, and a {\em cocoon} consisting
of shocked jet and shocked ambient material.  These are familiar
components from the theory of radio galaxies
\citep{1989ApJ...345L..21B}; see figure \ref{fig:diagram}. 

\subsection{Presence of an uncollapsed envelope} 
\label{S:freefalltimes}
\begin{figure*}
\centerline{\epsfig{figure=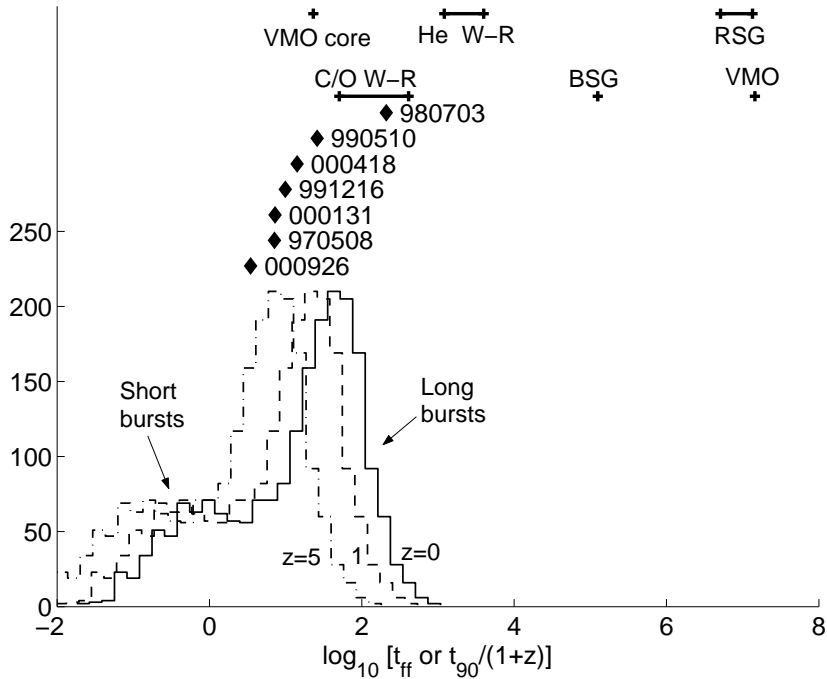, height=3.6in}} 
\caption[GRB durations compared to stellar dynamical times.]
{\footnotesize GRB durations compared to stellar dynamical times.  The
histogram of observed burst durations ($t_{90}/(1+z)$, for redshifts $z=0$,
$z=1$ and $z=5$) from the current BATSE catalog
(http://cossc.gsfc.nasa.gov/batse/) is plotted to show the timescales
for long and short bursts.  Notable bursts with afterglow observations
are plotted as diamonds and labeled. Also plotted are the typical
free-fall times for carbon-oxygen (``WR: C/O'') and helium (``WR:
He'') progenitors evolved from Wolf-Rayets, blue and red supergiants
(``BSG'' and ``RSG,'' respectively), the primordial $300~\Msun$
progenitor (``VMO'') calculated by \cite{2001ApJ...550..372F}, and the
$100 \Msun$ core of a VMO with no envelope.  For GRBs arising from
SNe, this plot shows that an uncollapsed stellar envelope remains to
be traversed by the jet -- except in the case of especially long GRBs
coming from compact carbon-oxygen stars or VMO cores.
\label{fig2} }
\end{figure*}

Stellar envelopes pose a problem for the propagation of GRB jets only
if they have not collapsed prior to the launching of the jet.  The
stellar envelope's collapse timescale is greater (probably by a factor
of at least a few) than its free-fall time $\tff$.  In figure
\ref{fig2} I compare stellar free-fall times with the intrinsic
durations of GRBs for several possible redshifts.  Only the very
densest progenitors, the cores of very massive objects ($\tff = 23
(M_{\rm core}/[100~\Msun])^{1/4}$ s at oxygen ignition) and helium
depleted post-Wolf-Rayet stars ($\tff \gtrsim 50$ s) could plausibly
collapse entirely in the durations of the longest GRBs, and then only
at low redshift.  In all other cases a stellar envelope remains to
impede GRB jets.

\subsection{Head Motion}\label{vhead}
For stellar core collapse to successfully produce a cosmological GRB,
it must emit a jet that clears the stellar envelope from the
observer's line of sight to the core.  This is required if the jet is
to achieve high Lorentz factors and if its internal shocks are to be
unobscured by overlying material.  This cannot occur if the Lorentz
factor of the jet head, $\Gamma_h\equiv (1-\beta_h^2)^{-1/2}$ (where
$\beta_h c$ is the head's velocity), exceeds the inverse beaming angle
$1/\theta$. If it did, then the jet head would be causally disconnected from
its edges and would behave like a spherical blastwave
\citep{1976PhFl...19.1130B}.  In this case essentially no material
would escape sideways to form the cocoon.  If instead $\Gamma_h<1/\theta$
then there is ample opportunity for shocked jet and envelope material
to flow sideways and inflate the cocoon, so a successful GRB requires
\begin{eqnarray}\label{GamThetaEnv}
\Gamma_h < 1/\theta & ({\rm stellar~envelope}). 
\end{eqnarray} 
Observers of GRB afterglows often identify an achromatic break in the
light curve with the deceleration of the swept up shell from $\Gamma_h
>1/\theta$ to $\Gamma_h<1/\theta$ \citep{1999ApJ...525..737R}.  In
order for this to be possible, the head must have satisfied 
\begin{eqnarray}\label{GamThetaWind}
\Gamma_h > 1/\theta & (\rm circumstellar~medium)
\end{eqnarray}
while it was still being driven forward by the jet (prior to its 
deceleration).  Condition (\ref{GamThetaWind}) only applies to GRBs
whose afterglows exhibit such a break in their afterglows. 

The expansion velocity of the jet head is given to a very good
approximation by the balance of jet and ambient ram pressures (momentum
fluxes) in the frame of the jet head.  This approximation is most accurate when
ambient material is cast aside into the cocoon, as is the case if
equation (\ref{GamThetaEnv}) is satisfied. 
Ram pressure balance means 
\begin{equation}
\rho_j h_j (\Gamma\beta)_{jh}^2 c^2 + p_j = 
\rho_a h_a (\Gamma\beta)_h^2 c^2+ p_a
\end{equation}
where $(\Gamma\beta)_{jh}$ is the relative four-velocity between the
jet and its head, $(\Gamma\beta)_{h}$ is the four-velocity of the head
into the ambient medium, $\rho$, $p$, and $h\equiv (e + p)/(\rho c^2)$ are
density, pressure, and enthalpy ($e$ is total comoving energy density),
and the subscripts $j$ and $a$ refer to jet and ambient material.  As
$p_a \ll \rho_a c^2$ in a stellar envelope $h_a=1$ to a good approximation
and $p_a$ can be ignored on the right-hand side.  To leading order in
$\Gamma_j^{-1}$, $p_j$ can be ignored on the left.  For a
stationary ambient medium, these approximations give
\begin{equation}\label{vHead}
\beta_h = \frac{\beta_j}{1 + \Lt^{-1/2} }
\end{equation}
\citep{1994A&A...281L...9M} where 
\begin{eqnarray}\label{LtildeDef}
\Lt &\equiv& \frac{\rho_j h_j \Gamma_j^2}{\rhoa} =
\frac{\Liso/c^3}{d\Ma/dr}. 
\end{eqnarray}
The second equality, which holds to leading order in $\Gamma_j^{-1}$,
is derived by noting that the kinetic plus internal energy density of
the jet, evaluated in the 
lab frame, is $\Gamma_j^2 h_j \rho_j -\Gamma_j \rho_j -p_j$; its
isotropic luminosity $\Liso$  is this quantity times $4\pi
r^2 \beta_j$; and the mass per unit length of the
ambient medium is $d\Ma/dr\equiv 4\pi r^2 \rhoa$.  Here $r$ is the
radius and the true jet luminosity is $L = (\theta^2/4)\Liso$.

The relation between $\Gamma_h$ and $\Gamma_j$ implied by equation
(\ref{vHead}) takes simple limits in two regimes.  If $\Lt \gg
\Gamma_j^4$, then the reverse shock into the jet is nonrelativistic
\citep{1995ApJ...455L.143S} and $\Gamma_j - \Gamma_h\ll \Gamma_j$.
This case is physically unattainable within a star and is achieved
outside only if the ambient density is extremely low.  In the 
opposite limit of a relativistic reverse shock, 
\begin{equation}\label{LimitsOfGammah} \Lt^{1/2} =
\frac{1}{\beta_h} -1  = 
\left\{
\begin{array}{lc} 
2\Gamma_h^2, & \Gamma_j^4 \gg \Lt \gg 1;  
\\ \beta_h, & \Lt \ll 1  
\end{array}\right. 
\end{equation} 
\citep[see also][]{2001PhRvL..87q1102M}. 
Therefore, $\Lt$ is useful in determining the observer's time 
$\tobs/(1+z) = t-r/c$ for the jet head to expand to radius $r$ if
viewed at redshift $z$.   Since
$d t = d r/(\beta c)$, 
\begin{equation}\label{tobs}
\frac{d \tobs}{1+z} = \frac{d r}{\Lt^{1/2} c}
\end{equation}
for both non-relativistic and relativistic jet heads, so long as
$\Lt\ll \Gamma_j^4$.  In presupernova envelopes and stellar winds
$d\Ma/dr$ is relatively constant, and since $\Liso$ is likely to vary
slowly $\Lt$ can be approximated with its average value.  More
generally, one might know how $\Liso$ varies as a function of $t$
(which is also $\tobs/(1+z)$ at the origin, $r=0$).  If one also knows
$\Ma(r)$ (e.g, from a model of the star and its collapse), then
equation (\ref{tobs}) integrates to
\begin{equation}\label{IntegratedTobs}
\int^{\tobs} \Liso(\tobs)^{1/2} \frac{dt'}{1+z} = \int^r \left(c
\frac{d\Ma}{dr}\right)^{1/2} dr',
\end{equation}
giving $\tobs(r)$ implicitly. 

\subsection{Cocoon Structure}\label{cocoon}
The jet cocoon is the region containing spent jet material and shocked
ambient material. Its extent is equal to that of the jet, but its
width is determined either by pressure balance with the surrounding
gas, if there is time for this to be achieved, or else by the
expansion of a lateral shock into the envelope.  The latter case holds
so long as it predicts a cocoon pressure $p_c$ in excess of the
hydrostatic pressure or collapse ram pressure $p_a$ in its
environment.  Let us first consider the case of an adiabatic cocoon
whose pressure exceeds that of its surroundings, and check this
assumption in \S \ref{S:PressureConf}.  Let us also restrict attention
to the case where the head velocity is subrelativistic ($\Lt <1$,
eq. [\ref{LimitsOfGammah}]), in which case the cocoon pressure $p_c$
is roughly constant away from the jet head.

The cocoon created by a jet of constant opening angle expands
self-similarly so long as its width $R_c$ expands in proportion to its
extent $r$, and so long as no other size scales affect its structure.
Under these conditions, numerical simulations can determine cocoon
structures exactly.  However, analytical estimates
\citep[e.g.,][]{1989ApJ...345L..21B}, while not as accurate, elucidate
how cocoon properties scale with $\Lt$ and $\theta$.

The cocoon expands nonrelativistically in a direction normal to its
surface at the velocity $\beta_c$ given by 
\begin{equation}\label{betacPrelim}
\rhoa c^2 \beta_c^2 = p_c. 
\end{equation}
If $\rhoa$ is evaluated at the point where the cocoon is widest, the
normal direction is sideways and thus $\beta_c \simeq R_c/(ct)$.  The
pressure $p_c$ is related to the energy $\Ein$ deposited in the cocoon
and the cocoon volume $V_c$ through $p_c \simeq \Ein/({3\rhoa
V_c})$ (an overestimate, as part of $\Ein$ is kinetic). 
If a cocoon of length $r$ and width $R_c$ is idealized as a cone, $V_c
\simeq (\pi/3) R_c^2 r$ and $\rhoa \simeq \rhoa(r)$ in equation
(\ref{betacPrelim}). 
But, since $\dot{r} = \beta_h c$ and $\dot{R}_c
\simeq \beta_c c$, $V_c \simeq (\pi/3) r^3 (\beta_c/\beta_h)^2$. 

Now, $\Ein$ is the energy emitted in time to catch up with the jet
head at radius $r$.  (After breakout, $\Ein$ is available to drive an
observable outflow and possibly a precursor: see
\cite{2002MNRAS.337.1349R} and \S \ref{Breakout}.)  As long as $\Lt\ll
\Gamma_j^4$ the flight time of the jet can be neglected compared to
$\tobs(r)/(1+z)$, so $\Ein(r)$ is essentially all of the energy
emitted up to then:
\begin{eqnarray}\label{Ein-from-L-tobs}
\Ein(r) &=& \int_0^{\tobs(r)/(1+z)} L~ dt \nonumber \\ 
&\simeq& \left[\frac{\theta^4}{16} \Liso r M(r) c  \right]^{1/2} 
\end{eqnarray}
using equation (\ref{tobs}).  Along with the expressions for $p_c$,
$V_c$ and $\beta_c$, this implies
\begin{eqnarray}\label{betacSoln}
{\beta_c} \simeq \Lt^{3/8} \theta^{1/2}. 
\end{eqnarray}
This implies that the cocoon is nonrelativistic (and slower than the
jet head) so long as the jet head is also nonrelativistic, for $\Lt < 1$
in this case (eq. [\ref{LimitsOfGammah}]), and $\theta < 1$ for a
collimated jet.  

\subsubsection{Cocoons vs. Jet-Driven Explosions}

Equation (\ref{betacSoln}) assumes that the length of the cocoon is
set by the advance of the jet head; this requires 
$\beta_c<\beta_h$. 
If this condition is violated, the cocoon expands around the jet and
develops into a roughly spherical blastwave \citep[see
also][]{1989ApJ...345L..21B}.  For a nonrelativistic jet head,
equations (\ref{LimitsOfGammah}) and (\ref{betacSoln}) indicate that
$\beta_c<\beta_h$ so long as $\Lt \gtrsim \theta^4$.  For a more
precise criterion, consider the velocity $\beta_{\rm bw}$ of a
spherical blastwave powered by two jets: 
\begin{equation}\label{betaBW}
\beta_{\rm bw} = 0.90 \left[\frac{2\Ein(t=r/\beta_{\rm bw})}{\Ma(r)
c^2}\right]^{1/2} = 0.74  \theta^{2/3} \Lt^{1/3},
\end{equation}
where the coefficient is given by the PGA/$\overline{K}$ approximation
of \cite{OM88}, for a wind bubble with a ratio of specific heats
$\hat{\gamma}=4/3$ in a medium with $\rhoa\propto r^{-2}$. The head
outruns this blastwave ($\beta_h>\beta_{\rm bw}$) if $\Lt >
({\theta}/{90^\circ})^4$, i.e., 
\begin{equation}\label{L:No-BW:L<1}
\frac{\Liso r}{c} > \left(\frac{\theta}{90^\circ}\right)^4 M(r) c^2. 
\end{equation}
This condition can only be violated if $\Lt <
(\theta/90^\circ)^4\ll1$, so that only jets driving nonrelativistic
heads can violate it.  I assumed a nonrelativistic head in deriving
equation (\ref{L:No-BW:L<1}); however there is nothing to suggest that
relativistic jet heads can be swallowed by their cocoons.

A successful GRB requires that stellar envelope material be cleared
from the path of the jet so it can sustain $\Gamma_j\gtrsim 10^2$ at
$r\sim$ 1 AU where internal shocks are thought to form.  A jet-cocoon
structure is required for this: a spherical blastwave would not
accomplish it.  For this reason GRBs must satisfy
$(\theta/90^\circ)^{4}<\Lt<\theta^{-4}$ within the stellar envelope
(eqs. [\ref{GamThetaEnv}],~[\ref{LimitsOfGammah}], and [\ref{L:No-BW:L<1}])
so that a cocoon forms but does not overcome the jet.  Of these limits
on $\Lt$, only the lower need be considered because $\Lt \ll
\theta^{-4}$ for any reasonable combination of opening angle and
stellar model.

One can use equation (\ref{L:No-BW:L<1}) in equation
(\ref{Ein-from-L-tobs}) to eliminate either $\Liso r/c$, which gives a
lower limit on $\Ein(r)$, or to eliminate $\theta^4 M(r) c^2$, which
gives an upper limit on $\Ein(r)$.  In terms of the total energy per
lobe $\Ein(R)$ deposited in the stellar envelope, the formation of a
cocoon rather than a spherical blastwave implies 
\begin{equation}\label{EinConstraint}
0.10\, {\theta^4} \Menv c^2 < \Ein(R) < 0.61 \Liso R/c. 
\end{equation}
A spherical blastwave is not sensitive to the site of energy
injection, whether at a jet head or near the collapsing core, and for
this reason the upper bound applies to energy from {\em any} source
that is entrained in the stellar envelope before the jet breaks out of
the star.  Note that the lower bound implies that $\Ein(R)$ must be at
least a fraction $(\theta/37^\circ)^2$ of the rest energy
of the envelope in the jet's path, $(\theta^2/4)\Menv c^2$. 

\subsection{Timing Constraint}\label{SS:timingconstraint}

One further and very important constraint derives from the durations
of GRBs.  In the internal-shock model, fluctuating gamma-ray emission
reflects variability in the central engine and persists only while
this engine is running \citep{1995ApJ...455L.143S}.  Define $\tgam$ as
the observed duration of a GRB, including any precursor but excluding
its afterglow.  In the collapsar model, this must be preceded by the
cocoon phase during which the jet crosses the stellar envelope.  The
central engine must therefore be active for at least $\tgam +
\tobs(R)$ in the observer's frame.  It is unlikely -- though not
impossible -- for $\tgam$ to be much shorter than $\tobs(R)$, for this
would require the central engine to shut off just as its effects
become observable. So
\begin{eqnarray}\label{tobs-tgamma-Rstar}
\frac{\tobs(R)}{1+z}\lesssim t_\gamma; &~&
\varepsilon_\gamma \Ein(R)\lesssim E_\gamma,
\end{eqnarray}
where $\varepsilon_\gamma$ is the efficiency with which the jet's kinetic
luminosity is converted into gamma rays in the observed band.  The
second inequality derives from the first as long as the jet's
luminosity is relatively constant.

\subsection{Jet variability}\label{S:Variability}
I have assumed a constant luminosity jet up to this point, but GRBs
are often observed to fluctuate significantly in intensity on very
short timescales.  How should the above results be adjusted for jet
variability?  First, note that the process producing gamma rays
(e.g., internal shocks) is likely to accentuate the intrinsic
variability of the source.  Second, the observed propagation speed of
the jet head, $dr/dt_{\rm obs}$, is equal to $c \Lt^{1/2}$
(eq. [\ref{tobs}]).  A fluctuating jet thus progresses more slowly than a
steady jet of the same mean luminosity.  If one uses
the average value of $\Lt$ to constrain a star's mass and radius by
requiring that $\tobs(R)$ is not long compared to the observed burst
(eqs. [\ref{tobs-tgamma-Rstar}] above and [\ref{r-M-L-tobs}] below),
then this constraint is only tightened if one accounts for
variability. Similarly, a variable jet deposits more energy per radius
than a steady jet of the same mean luminosity; this only tightens the
constraints derived by requiring a jet-cocoon structure
(eq. [\ref{L:No-BW:L<1}]).

\section{Constraints on Stellar Hosts}\label{EnvelopeConstraints}
To apply the above constraints to stellar progenitors for GRBs,
observed quantities must be related to the the parameters of equations
(\ref{L:No-BW:L<1}), (\ref{EinConstraint}), and
(\ref{tobs-tgamma-Rstar}).  This is possible for cases where, in
addition to $\tgam$, the redshift $z$ and jet opening angle $\theta$
have been derived from afterglow observations.  To construct $\Lt$ one
requires the isotropic (equivalent) kinetic luminosity $\Liso$ and the
mass per unit radius $d\Ma/dr$ in the environment.

I adopt for GRBs' energy, luminosity, and duration the following
definitions: $\Lisogam \equiv \Eisogam(1+z)/ \tgam$; and
$\{\Egam,\Lgam\} \equiv \{\Eisogam, \Lisogam \} \times \theta^2/4$.
Here the isotropic gamma ray energy is $\Eisogam$, and
$\varepsilon_\gamma$ is the average efficiency factor relating the
gamma-ray energy to the total (kinetic, Poynting, and photon)
luminosity of the jet.  The net energy $\Etot$ and luminosity $\Ltot$
represent only the approaching jet, which presumably has a counterjet.

A key assumption employed throughout this section is that the
$\gamma$-ray half opening angle $\theta$ is equal to the jet's
half opening angle  while it crosses the outer stellar
envelope.  This is essentially the same as the assumption that the jet
is ballistic (rather than hot and pressure-confined) at that point;
this assumption is revisited in \S \ref{Widening}.
I assume the value of $\theta$ derived from afterglow observations
\citep{1999ApJ...525..737R} can be used to characterize the jet as it
crosses the outer stellar envelope and emerges from the star.
The above definitions identify $\Lisogam$ as the mean value (energy
$\Egam$ in source-frame duration $\tgam/(1+z)$).  This is somewhat
arbitrary, since GRBs are highly variable; see \S \ref{S:Variability}
for justification.

For numerical evaluations I use either $\Lisofto \equiv
\Lisogam/10^{51}$ erg/s, or $\Efpf\equiv \Egam/10^{50.5}$ erg.  The
former is a characteristic value for GRBs and can be observed without
determining $\theta$.  The latter is motivated by
\cite{2001ApJ...562L..55F}'s result that $\Egam = 10^{50.5\pm
0.5}~\erg$ for ten GRBs whose $\theta$ could be determined.

In a stellar envelope the radial average value of $d\Ma/dr$ is
$\Menv/R$, the ratio of envelope mass to stellar radius.  Indeed, many
presupernova envelopes have density profiles close to $\rhoa\propto
r^{-2}$ (\citealt{1989ApJ...346..847C}), for which
$d\Ma/dr$ is constant at its average value.  The
average value of $\Lt$ is 
\begin{eqnarray}\label{LtildeEnv}
\Lt = 1.30\times 10^{-3}\frac{\Lisofto}{\varepsilon_\gamma}
\frac{R}{\Rsun}\frac{\Msun}{\Menv} & ({\rm envelope}). 
\end{eqnarray}
In order for the jet head to be relativistic ($\Lt>1$), the star
must have a mass per length much lower than $\Msun/\Rsun$, or the
gamma-ray efficiency $\varepsilon_\gamma$ must be small.

\subsection{Constraints from Burst Duration and Cocoon Formation}
Under the assumption of a ballistic jet, the observed time of jet
breakout is
\begin{equation}\label{tobs-R-evaluated}
\tobs(R) = 64.4(1+z) \left(\frac{\varepsilon_\gamma}{\Lisofto}
\frac{R}{\Rsun} \frac{\Menv}{\Msun}\right)^{1/2}
~\s. 
\end{equation}
This, along with the constraint $\tobs(R)\lesssim \tgam$
(eq. [\ref{tobs-tgamma-Rstar}]) illustrates that typical long-duration
GRBs are most easily produced in compact stars
\citep{2001ApJ...550..410M}.  This constraint is best expressed
\begin{equation}\label{r-M-L-tobs}
\varepsilon_\gamma \frac{\Menv}{\Msun}\frac{R}{\Rsun} \lesssim  \Lisofto
\left[\frac{\tgam/(1+z)}{64.4  ~\s}\right]^2; 
\end{equation}
the left-hand side pertains to a hypothetical stellar model and to the
efficiency of gamma radiation and is constrained by observables on the
right.  Note that the combination $\Lisofto [\tgam/(1+z)]^2$ is
related to an observed burst's fluence and duration through its
comoving distance (i.e., physical distance at redshift zero), rather than
its luminosity distance.  Note also that the above constraint is
independent of $\theta$.

\begin{figure*}
\centerline{\epsfig{figure=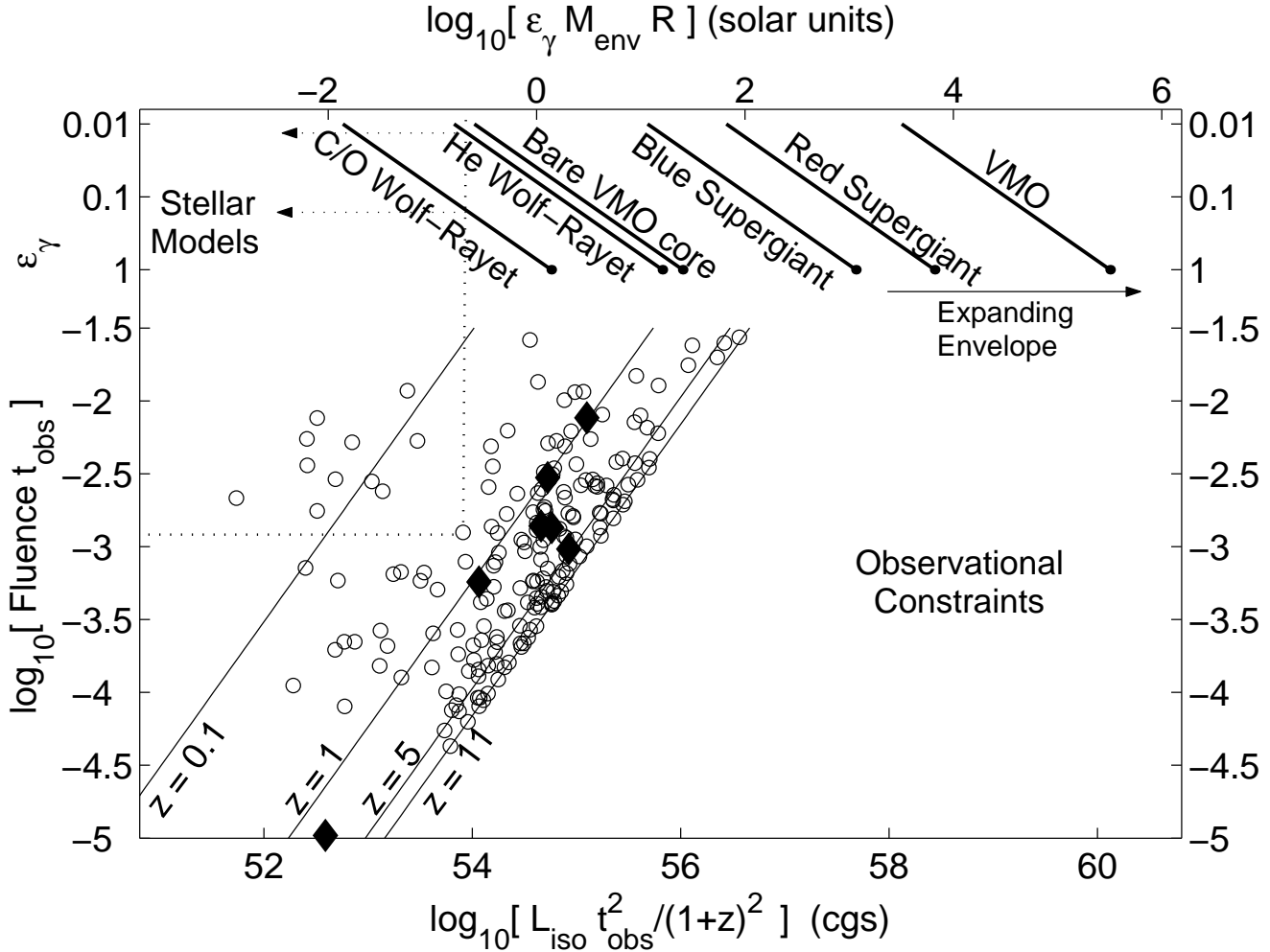 ,height=5.5in}}
\caption[Timing constraint.]  {\footnotesize The product of a GRB's
fluence and duration constrains the product of mass and radius in any
stellar envelope through which the GRB jet (if moving ballistically:
see \S \ref{Widening}) could have emerged rapidly compared to the
period of emission. The observed quantity fluence$\cdot t_{obs}$ and
the redshift imply the quantity $\Liso \tobs^2/(1+z)^2$ intrinsic to
the burst. By eq. (\ref{r-M-L-tobs}), this places an upper limit on
the model parameter $\varepsilon_\gamma \Menv R$, where
$\varepsilon_\gamma$ is the efficiency of gamma-ray production; so, a
model should lie to the left of its burst on the plot. This limit
insensitive to GRB redshift when $z>1$. In this plot {\em solid
diamonds} are the bursts plotted in figure \ref{fig2}, with redshifts
and fluences derived from \cite{2001ApJ...562L..55F} and references
therein; {\em circles} are 220 bursts for which \cite{GRBcepheid}
estimate redshifts using a luminosity-variability correlation (these
authors did not allow $z>11$). The burst illustrated with {\em dotted
lines} is compatible with a carbon-oxygen post-Wolf-Rayet progenitor
if $\varepsilon_\gamma\lesssim 15\%$, or with a helium-bearing
post-Wolf-Rayet star or the 100 $\Msun$ bare core of a very massive
object (VMO) if $\varepsilon_\gamma \lesssim 3\%$. (For other VMO core
masses, use $\Menv R \propto M^{3/2}$ at O ignition.) It is not
compatible with blue or red supergiant progenitors (BSGs or RSGs),
VMOs that have not lost their radiative envelopes, or the expanding
envelope of the ``supranova'' progenitor of \cite{1999ApJ...527L..43V}
(not shown), which moves to the right on this plot.  Similar
conclusions can be drawn for almost all of the bursts plotted.  Lines
of constant $z$ are shown for $\Lambda=0.7$, $\Omega_m=0.3$, $H_0 =
65~\kps~\Mpc^{-1}$.
\label{figTiming} }
\end{figure*}

In figure \ref{figTiming} I illustrate the derivation of
$\Liso\tobs^2/(1+z)^2$ and constraint (\ref{r-M-L-tobs}) from the
observed fluence and duration and an observed or estimated redshift
(for $\Omega_m = 0.3$, $\Lambda=0.7$, $H_0 = 65\ \kps\Mpc^{-1}$).
Bursts with well-observed redshifts are plotted as solid diamonds;
those for which \cite{GRBcepheid} have estimated redshifts from a
luminosity-variability correlation are plotted as open circles.  
A progenitor model passes the timing constraint if it lies to the left
of its burst. 
Most of the bursts plotted are consistent with $\varepsilon_\gamma
\Menv R\lesssim 10 \Msun \Rsun$; for none of them does this limit
exceed $90 \Msun\Rsun$.  This constraint is nearly independent of
redshift and derives primarily from the distribution of $({\rm
fluence}\cdot \tobs)$ among GRBs.  VMO cores and Wolf-Rayet stars are
compatible with many bursts. Blue supergiants are compatible only with
the very brightest and longest. Red supergiants and VMOs that retain
their envelopes are ruled out, as are optically-thick shrouds of
expanding ejecta which might persist in the ``supranova'' or
stellar-merger models of central engines.

\begin{figure*}
\centerline{\epsfig{figure=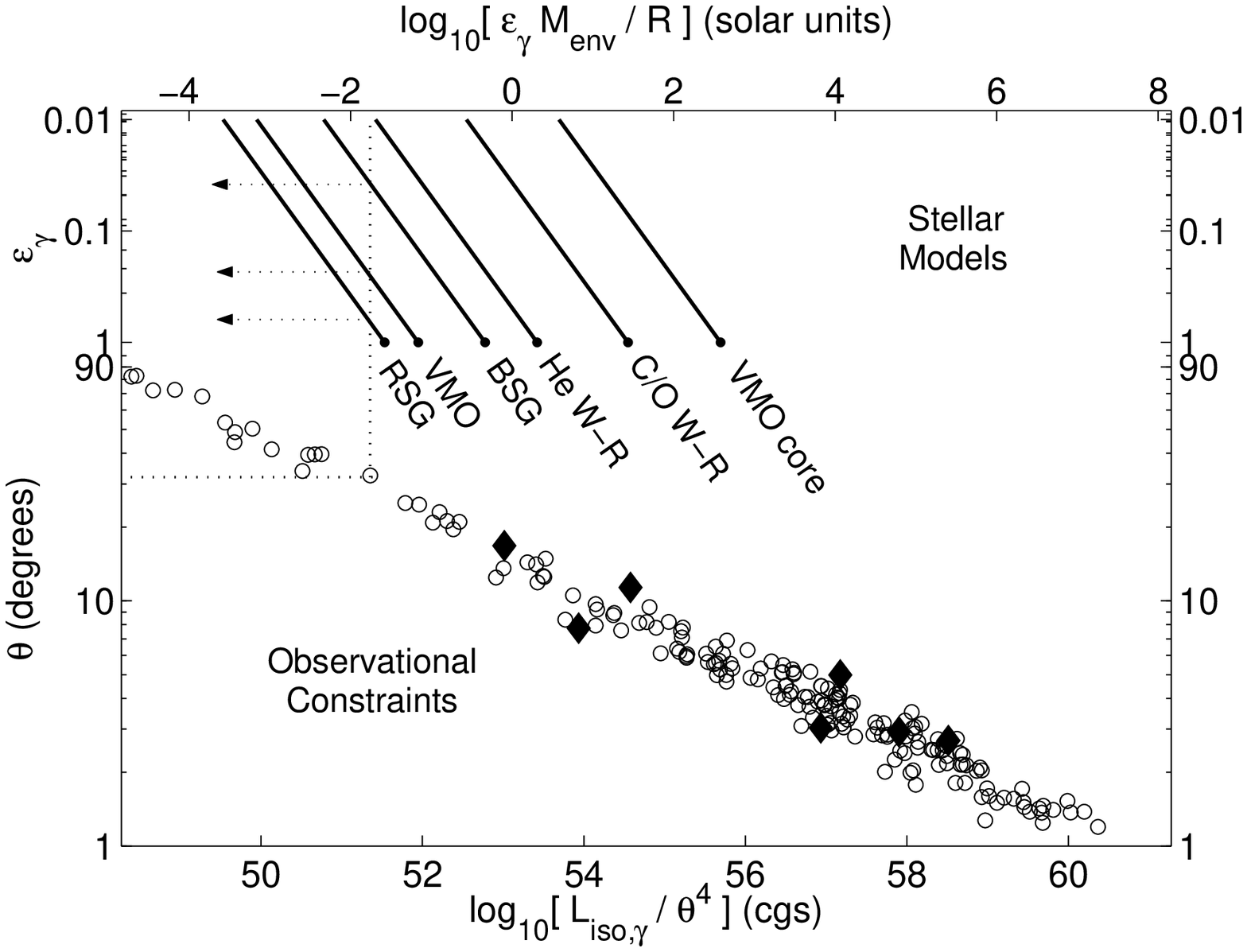 ,height=4in}}
\caption[Cocoon constraint.]  {\footnotesize The opening angle and
isotropic luminosity of a GRB -- in the combination $\Liso
\theta^{-4}$, eq. (\ref{r/eps*M-theta-tobs}) -- constrain the ratio of
mass and radius in any envelope through which the GRB's jet can travel
without being engulfed within its own bubble. (The jet is assumed to
be moving ballistically; see \S \ref{Widening}.) As in figure
\ref{figTiming}, a viable model should lie to the left of its GRB on
the plot.  Here, the constraint is most sensitive to $\theta$.
{\em Filled diamonds} represent bursts with $\theta$ inferred from
afterglow observations \citep[][and references
therein]{2001ApJ...562L..55F}.  {\em Open circles} are the bursts for
which \cite{GRBcepheid} estimate redshifts; for these $\theta$ is
estimated under the assumption of a common energy $E_\gamma =
10^{50.5}$ erg (causing $\theta$ to correlate tightly with
$\Lisogam$).
Most bursts are compatible with all stellar progenitors by this
criterion; however, those with large $\theta$ are not. Diffuse stars
most easily pass this test. VMO cores fail first, when $\theta\simeq
10^\circ$ (plotted is a $100~\Msun$ core; $\Menv/R \propto M^{1/2}$ at
O ignition). The burst illustrated by {\em dotted lines} is compatible
with RSGs, VMOs that retain their envelopes, BSGs, and helium
post-Wolf-Rayet stars for values of $\varepsilon_\gamma$ less than
$90\%, 50\%, 3\%$, and $1\%$, respectively. The supranova model
travels to the left on this plot as it expands, and easily passes this
constraint.
\label{figCocoon} }
\end{figure*}

Equation (\ref{L:No-BW:L<1}) gives a complimentary constraint
on the basis that a jet-cocoon structure exists: 
\begin{equation}\label{r/eps*M-theta-tobs}
{\varepsilon_\gamma} \frac{\Menv}{\Msun} \frac{\Rsun}{R} < {\Lisofto}
\left(\frac{17.0^\circ}{\theta}\right)^4. 
\end{equation}
This constraint is derived from observations of bursts in figure
\ref{figCocoon}. Similar to figure \ref{figTiming}, {filled diamonds}
here represent bursts for which afterglow observations have allowed an
estimate of $\theta$ in addition to $\Liso$ \citep[as reported by][and
references therein]{2001ApJ...562L..55F}; open circles represent redshifts
estimated by \cite{GRBcepheid}.  For these I have estimated $\theta$
by requiring that $E_\gamma = 10^{50.5}$ erg as suggested by
\cite{2001ApJ...562L..55F}.  These points fall in a narrow band on the
plot because $\Lisogam/\theta^4 =(1+z) E_\gamma /(\theta^4 \tobs)$.
With $E_\gamma$ held fixed, the dispersion in this quantity is
dominated by $\theta^{-4}$ (2 dex rms, for the points plotted) rather
than $(1+z)/\tobs$ (0.5 dex rms).

\section{Constraints on Stellar Winds}\label{WindConstraints}
GRBs from supernovae are likely to occur within a dense stellar wind,
an environment that potentially affects both the GRB itself and its
afterglow \citep{2000ApJ...541L...9K, 2001MNRAS.327..829R}.

The internal-shock model for GRB emission posits that significant
variability on a time scale $\delta t$ ($\equiv \delta \tobs/(1+z)$) arises
from the collision of shells within the jet at radii of about
\begin{equation} \label{rIS}
\rIS \equiv 2\Gamma_j^2 c \; \delta t, 
\end{equation}
as discussed by \cite{1994ApJ...430L..93R}. \cite{1995ApJ...455L.143S}
argue that an external shock with the ambient medium -- to which the
afterglow is attributed -- cannot create a fluctuating gamma ray
burst.  For internal shocks to occur, the external shock (jet head)
must move beyond $rIS$ on a timescale not long compared to the
duration of the burst: $\Delta\tobs(\rIS) \lesssim \tgam$.

\cite{1995ApJ...455L.143S} define $N_p\equiv \tgam/\delta \tobs$ as the
number of pulses that fit within the burst duration given a
characteristic separation $\delta \tobs$.  With equations
(\ref{LimitsOfGammah}), (\ref{rIS}), and this definition, the
criterion $\Delta\tobs(\rIS) \lesssim \tgam$ becomes 
\begin{equation}\label{GhGj-head-past-rIS}
\Gamma_h^2 \gtrsim \frac{\Gamma_j^2}{N_p}, 
\end{equation} 
or 
\begin{equation}\label{Lt-head-past-rIS}
\Lt \gtrsim \frac{4\Gamma_j^4}{N_p^2} = 4\times 10^4
\left(\frac{\Gamma_j}{100}\right)^4 
\left(\frac{100}{N_p}\right)^{2}. 
\end{equation}
\citeauthor{1995ApJ...455L.143S} argue that $N_p\sim 100$ is typical,
although a wide variety of time scales is observed within bursts
\citep{2000ApJS..131....1L}.  

For a wind $d\Ma/dr$ is the ratio of mass loss rate to wind velocity,
$\dot{M}_w/v_w$, so
\begin{eqnarray}\label{LtildeWind}
\Lt = \frac{5.89\times 10^6}{\varepsilon_\gamma} \frac{\Lisofto v_{w,8}}
{\dot{M}_{w,-5}} & ({\rm wind}) 
\end{eqnarray}
where $v_{w,8} \equiv v_w/(10^8~\cm~\s^{-1})$ and ${\dot{M}_{w,-5}}=
\dot{M}_w/(10^{-5} ~\Msun~\yr^{-1})$ are normalized to characteristic
values for Wolf-Rayet stars \citep{2000A&A...360..227N}, although the
presupernova mass loss rates of such stars is not well known
\citep{1991supe.conf..549L}.  Expressing condition
(\ref{Lt-head-past-rIS}) in terms of $\Gamma_j$,
\begin{equation}\label{GjMax}
\Gamma_j <  350 \left(\frac{N_p}{100}\right)^{1/2}
\left(\frac{\Lisofto v_{w,8}}
{ \varepsilon_\gamma\dot{M}_{w,-5}}\right)^{1/4}.  
\end{equation}
The analogous criterion for a uniform ambient medium was presented by
\cite{1999PhR...314..575P}.  This upper limit on
$\Gamma_j$ must exceed the lower limit required for the escape of the
observed gamma rays, described most recently by
\cite{2001ApJ...555..540L}.  These authors inferred $\Gamma_j>340$ for
five of the ten ordinary bursts in their Table 2; equation
(\ref{GjMax}) suggests that these could not have occurred in a stellar
wind representative of Wolf-Rayet stars -- unless, for instance,
$N_p\gg 100$ (which does not appear to characterize the bursts in
their Table 2).   

However, condition (\ref{GjMax}) might be circumvented by the
efficient sweeping-forward of the ambient medium by runaway pair
production, as discussed by \cite{2000ApJ...534..239M},
\cite{2000ApJ...538..105T} and \cite{beloborodov-sweeping}.  If the
relativistic flow has generated a fraction of the observed photons at
a radius smaller than $\rIS$, then these can clear optically-thin
ambient gas from the region. 

A spreading break 
\citep[albeit one difficult to detect:][]{2000ApJ...541L...9K}
will occur in the afterglow if
equation (\ref{GamThetaWind}) is satisfied.  The remaining isotropic
kinetic luminosity after the gamma-rays have been emitted is
$(1-\varepsilon_\gamma) \Lisogam /\varepsilon_\gamma$.  Along with
(\ref{LtildeWind}), this becomes
\begin{equation}\label{LtWindCnstrnt-LimOnTheta}
\theta > 1.64^\circ \left[\frac{\varepsilon_\gamma\dot{M}_{w,-5}}
{(1-\varepsilon_\gamma)\Lisofto v_{w,8}} \right]^{1/4}. 
\end{equation}
This criterion could only be violated by the narrowest of GRB jets in
especially dense stellar winds.

\section{Breakout and Blowout: Observational Consequences} \label{Breakout}

The energy deposited by the jet as it crosses a surrounding stellar
envelope ($\Ein$ in \S \ref{Inside Star}) is available to power
(concievably) observable phenomena distinct from the GRB and its
afterglow.  These include transients associated with the emergence of
the jet head from the stellar surface ({\em breakout}), and the escape
of hot cocoon material along the path of the jet ({\em blowout}).
Note that a type Ic light curve is {not} a product of jet energy,
as SN Ic are powered by radioactive nickel, which is not produced at
the jet head.

The energy $\Ein$ is typically less than the total GRB jet
energy (for successful GRBs) according to condition
(\ref{tobs-tgamma-Rstar}).  Nevertheless the effects of deposited
energy may be of observational interest if they cover a wider opening
angle, dominate a different observed band, are emitted with higher 
efficiency, or occur where the jet stalls before breakout.

\subsection{Shock Acceleration and Breakout}\label{SS:Breakout}
In ordinary supernovae the fastest ejecta are produced at the surface
of the star.  There, the rapid decrease in stellar density leads to a
whip-like acceleration of the shock front followed by additional
postshock acceleration \citep{S60}.  Sufficiently energetic explosions
can produce relativistic ejecta this way, so long as the stellar
progenitor is sufficiently compact \citep{1999ApJ...510..379M}.  The
impact of these ejecta with a circumstellar wind may produce a
transient of hard photons; this is a plausible explanation for the
association between supernova 1998bw and GRB 980425
\citep{1999ApJ...510..379M,2001ApJ...551..946T,2001gra..conf..200M}. 
\footnote{SN 2003dh, which appeared in the afterglow of GRB 030329,
appears similar to SN 1998bw; however, the Tan et al. model cannot
explain a two-peak GRB of $\sim 10^{51.5}$ erg emerging from a SN of
comparable energy.}

\cite{1999ApJ...524..262M}, \cite{2001ApJ...550..410M}, and
\cite{2003ApJ...586..356Z} have
suggested that shock breakout might lead to the X-ray precursors seen
in some bursts \citep{1984ApJ...286..681L, 1991Natur.350..592M,
1999ApJ...516L..57I}.  \citeauthor{1999ApJ...524..262M} and
\citeauthor{2001ApJ...550..410M} concentrate on the prompt flash from
the shock-heated stellar photosphere; however, the fast ejecta have a
greater store of kinetic energy to be tapped.
\cite{2000ApJ...543L.129L} discuss several cases of precursor activity
in gamma rays, including one burst for which iron-line emission has
been suggested \citep[GRB 991216:][]{2001ApJ...550L..43V}.
If precursors are due to shock breakout then their energy is a small
fraction of $\Ein(R)$ and is therefore limited by conditions
(\ref{EinConstraint}) and (\ref{tobs-tgamma-Rstar}); however it may
yet be observable. 
\cite{1999ApJ...516L..57I} argue that the X-ray precursor of GRB
980519 resembles its X-ray afterglow to the extent that the afterglow
effectively preceded the GRB; as argued by \cite{1998ApJ...494L..45P},
this is consistent with the emission from material with
$\gamma<\Gamma_j$ ejected prior to the GRB emission.
It is therefore worthwhile to estimate the distribution of ejecta
kinetic energies from shock breakout in a jetlike explosion.

To do this one must identify at what point the jet head makes a
transition from the state of ram pressure balance described in \S
\ref{vhead} to the state of whip-like shock acceleration discussed by
\cite{2001ApJ...551..946T}.  The jet head obeys ram pressure balance
only if shocked ambient material exits the jet head more rapidly than
the head accelerates.  While this holds, the forward ambient shock
cannot travel far ahead of the jet reverse shock.  At some point near
the surface, however, ambient material cannot exit the jet head prior
to breakout.  The forward shock will then accelerate away down the
density gradient, and the flow becomes progressively more normal to
the surface \citep{2001ApJ...551..946T}.

If $\beta_\perp$ is a typical value for the perpendicular component of
velocity in the jet head, and $x \equiv (R-r)/R$ is the fractional
depth within the stellar envelope, then material is trapped within the
head if $(\theta R)/\beta_\perp > (x R)/\beta_h$, i.e., if the time to
exit the head exceeds the time to reach the surface.
A reasonable guess for $\beta_\perp$ is that shocked ambient gas exits
the jet head at the postshock sound speed, in the frame of the head.
For a nonrelativistic head, then, $\beta_\perp \simeq (8/49)^{1/2}
\beta_h$.  In the relativistic case, the transverse velocity saturates
at $\sim c/\sqrt{3}$ in the head's frame; in the star's frame,
$\beta_\perp \simeq 1/(3^{1/2} \Gamma_h)$ \citep{1999ApJ...525..737R}.
So the transition occurs when
\begin{equation}\label{xwhip}
x = \theta \times \left\{\begin{array}{cc} 
(49/8)^{1/2} \phi_{nr} , & \Lt(x) \ll 1; \\
3^{1/2} \Gamma_h\phi_r , & \Lt(x) \gg 1 \end{array}\right.
\end{equation}
where $(\phi_{nr},\phi_r)\sim 1$ are uncertain parameters.  In the
relativistic case, $x$ is determined implictly once $\Gamma_h(x)$ is
known. For this, apply equation (\ref{vHead}) to the outer density
distribution of the stellar progenitor 
\begin{equation}
\rhoa = \rho_h\left(\frac{x}{1-x}\right)^n,
\end{equation}
where $n$ is the effective polytropic index and the coefficient
$\rho_h$ is an extrapolation to $r=R/2$.  In a radiative outer layer
$\rho_h$ can be derived from the mass, radius, and luminosity of the
progenitor star \citep{1994sse..book.....K};
\cite{2001ApJ...551..946T} present formulae (their equations [25],
[45], and [48]) for $\rho_h$ in Kramers or Thomson atmospheres.

Once $x$ and $\beta_h(x)$ or $\Gamma_h(x)$ have been identified, the
production of fast ejecta follows from \cite{2001ApJ...551..946T}'s
theory (their eq. [54]) if one matches the shock four-velocity
$\gamma_s \beta_s$ with the head four-velocity $\Gamma_h\beta_h$ at the
depth $x$.   With equations (\ref{vHead}) and (\ref{xwhip}) setting a
reference depth, external mass, and shock velocity, their theory
predicts an isotropic-equivalent ejecta kinetic energy 
\begin{eqnarray}\label{EkIsoNR}
E_{k,{\rm iso}}(>\Gamma_f) &=&
\frac{(\phi_{nr} \theta)^{-1.65n\gamma_p} 
\Gamma_f^{-(1.58\gamma_p-1)}}
{(n+2.70)
\exp\left(1.98+7.03/n- 1.52n\right)}
\nonumber \\ &~&\times 
\left(\frac{\Liso}{R^2 \rho_h
c^3}\right)^{2.68\gamma_p-1}
\Liso \frac{R}{c}, 
\end{eqnarray}
when the transition occurs in the nonrelativistic regime, and 
\begin{eqnarray}\label{EkIsoXR}
E_{k,{\rm iso}}(>\Gamma_f) &=&
\frac{(\phi_{r} \theta)^{-\frac{0.317\gamma_p}{4\gamma_p-3}}
\Gamma_f^{-(1.58\gamma_p-1)} }
{(n+2.70)\exp\left[\frac{(n+3.26)(n+6.69)}{1.22n(n+4)}\right]} 
\nonumber \\ &~&\times 
\left(\frac{\Liso}{R^2 \rho_h
c^3}\right)^{\frac{4.32\gamma_p-3}{n(4\gamma_p-3)}}
\Liso \frac{R}{c}, 
\end{eqnarray}
when the head is relativistic at the transition.  Here,
$\gamma_p\equiv 1+1/n$ is the envelope's polytropic exponent.  Thesen
equations have been simplified by the restriction $\Gamma_f\gg 1$.

In general, the appropriate value of $E_{k,{\rm iso}}(>\Gamma_f)$ is
the minimum of the values given by equations (\ref{EkIsoNR}) and
(\ref{EkIsoXR}).  Note that $E_{k,{\rm iso}}(>\Gamma_f)$ is an
isotropic equivalent; the {\em total} kinetic energy per lobe in
ejecta above $\Gamma_f$ is smaller by $\theta^2/4$.  If one varies
$\theta$ holding the other quantities fixed, this {total} energy
above $\Gamma_f$ is maximized when the two expressions are equal, whereas
the {isotropic} value continues to rise slowly as $\theta$ is
decreased in the relativistic regime.  In the relativistic regime, the
above formulae only appy to $\Gamma_f > \Gamma_h(x)^{2.73}$ -- i.e.,
only to those ejecta involved in quasi-spherical shock acceleration.

The uncertainty of $(\phi_{nr},\phi_r)\times \theta$ leads to a much
greater uncertainty in $E_{k,{\rm iso}}(>\Gamma_f)$ for the
nonrelativistic than for the relativistic case: for instance, when
$n=3$, $E_{k,{\rm iso}}(>\Gamma_f)$ varies as
$(\phi_{nr}\theta)^{-6.7}$ in the nonrelativistic and as
$(\phi_{r}\theta)^{-0.19}$ in the relativistic regime.  Physically,
this difference arises because ram pressure and shock acceleration
give very different velocity laws in the nonrelativistic regime
($\beta_h\propto \rhoa^{-1/2}$ and $\beta_s\propto \rhoa^{-0.187}$,
respectively) whereas they give very similar velocity laws in the
relativistic regime ($\Gamma_h\propto \rhoa^{-1/4}$ and
$\Gamma_s\propto \rhoa^{-0.232}$, respectively).  Another uncertainty
concerns whether the head velocity should be matched to the shock or
postshock velocity; the latter choice increases the ejecta energies by
only $6-7\%$ for both relativistic and nonrelativistic transitions.

\subsubsection{An Example: the SN 1998bw Progenitor}\label{98bwBreakout}
To illustrate these estimates of jet-driven shock breakout, let us
consider the progenitor model for SN 1998bw adopted by \cite{WES98}
and studied by \cite{2001ApJ...551..946T}.  The parameters $n$, $R$,
and $\rho_h$ can be derived from \citeauthor{2001ApJ...551..946T}'s
tables 2 and 3.  Adopting their fit for the outermost regions (whose rest
energy is $<1.7\times 10^{50}$ erg), $n=4$, $\rho_h = 335~\g~\cm^{-3}$,
and $R = 1.4\times 10^{10}$ cm. The isotropic kinetic energy in
breakout ejecta is therefore:
\begin{eqnarray}\label{eq98bwBreakout}
E_{k,{\rm iso}}(>\Gamma_f) \simeq
\min 
\left[1.9 \left(\frac{\Lisofto}{\varepsilon_\gamma}\right)^{1.3}
\left(\frac{\phi_r\theta}{2.5^\circ}\right)^{-0.20},  
\right. \nonumber \\ \left.
      2.1\left(\frac{\Lisofto}{\varepsilon_\gamma}\right)^{3.3}
 \left(\frac{\phi_{nr}\theta}{2.5^\circ}\right)^{-8.4}
\right] 
\times 10^{48} \Gamma_f^{-0.975} ~\erg. 
\end{eqnarray}
If this formula were to give a result $>2\times 10^{50}$ erg,
a different envelope fit would be appropriate; however, the
qualitative result would be unchanged. 

Even in isotropic equivalent, the energy predicted by equation
(\ref{eq98bwBreakout}) is small compared to a cosmological GRB.  It is
comparable in magnitude to the values derived by
\citeauthor{2001ApJ...551..946T} for the spherical explosion of SN
1998bw.
While sufficient to produce GRB 980425 at a redshift of 0.0085, it
would not contribute to the appearance of a GRB at $z\gtrsim 1$.

It should not be surprising that the energy of motion in shock
breakout is intrinsically much smaller than that available in the jet,
as the accelerating shock is powered by the jet for a brief period (a
small range of radii) prior to breakout. Breakout does produce a spray
of ejecta with a variety of Lorentz factors, which may produce 
weak transients if observed off the jet axis. 

\subsection{Cocoon Blowout} \label{SS:Blowout}

The cocoon inflated by a jet prior to breakout is filled with hot gas
that is free to expand away from the star after breakout, constituting
a ``dirty'' fireball \citep{1998ApJ...494L..45P} which may be visible
either through its thermal emission, through its circumstellar
interaction, or through line absorption and fluorescence.  The cocoon
energy $\Ein$ (eq. \ref{Ein-from-L-tobs}) comprises
\begin{equation}\label{EinEvaluated}
4.4\times 10^{49} 
\left[
  \frac{\Lisofto}{\epsilon_\gamma}
  \left(\frac{\Menv}{\Msun}\right) 
  \left(\frac{R}{\Rsun}\right) 
\right]^{1/2} 
\left(\frac{\theta}{3^\circ}\right)^2
 {\rm erg}. 
\end{equation}

\cite{2002MNRAS.337.1349R} have considered observable implications of
cocoon blowout under the hypothesis that no mixing occurs between
shocked jet and shocked envelope material, so that the cocoon fireball
can accelerate to Lorentz factors comparable to $\Gamma_j$.  A likely
alternative is some mixing between jet and envelope, especially near
the jet head where shear is strong.  Suppose the shocked jet mixes
with the envelope material within an angle $\phi_{\rm mix} \theta$ of
the jet axis before entering the cocoon: $\phi_{\rm mix}\simeq 1$ for
mixing near the jet head only.  The energy per mass of the hot cocoon
material is then $(2/\phi_{\rm mix}) \Lt^{1/2} c^2$, corresponding to
a subrelativistic outflow velocity
\begin{eqnarray}\label{vBlowout}
v_{\rm out} &=& \frac{\sqrt{2}}{\phi_{\rm mix}} \Lt^{1/4} c\nonumber\\
&=& \frac{0.27}{\phi_{\rm mix}} 
\left[
\left( \frac{R}{\Rsun}\right) 
\left( \frac{\Msun}{\Menv}\right) 
\left( \frac{\Lisofto}{\epsilon_\gamma } \right)
\right]^{1/4} 
c
\end{eqnarray} 
where I assume $\Lt\ll 1$ as is appropriate for the compact
progenitors favored in \S \ref{EnvelopeConstraints}.  If blowout
occurs into an angle $\theta_{\rm out}$, then the expanding cocoon
material becomes optically thin ($\tau<1$) along its axis after roughly
\begin{eqnarray}\label{tThin}
&14& 
\left[
\left(\frac{\epsilon_\gamma}{\Lisofto}\right) \left(\frac{\Rsun}{R}\right)
\right]^{1/4}
\left(\frac{\Menv}{\Msun}\right)^{3/4} 
\nonumber \\ &&\times 
\left(\frac{\kappa}{0.4\,\cm^2\,\g^{-1} }\right)^{1/2}
\frac{\phi_{\rm mix} (\theta/3^\circ)}{\theta_{\rm out}} {~~~\rm
  hours}.
\end{eqnarray} 
Before this occurs, however, photons will diffuse sideways out of the
blowout cloud when the lateral optical depth $\tau_{\rm lat} \simeq
\theta_{\rm out}\tau$ and lateral velocity $v_{\rm lat}$ satisisfy
$\tau_{\rm lat} \simeq c/(3 v_{\rm lat})$.  This occurs at $\sim 13$
hours for the fiducial parameters listed above (independently of
$\theta_{\rm out}$); the amount of internal energy persisting in the
ejecta is quite small by that point, indicating that a thermal pulse
will not be observed unless the fireball is enriched with short-lived
isotopes.  The modest cocoon energy is available for circumstellar
interaction but may be masked by the GRB afterglow. The cocoon blowout
does, however, represent a screen between the jet-afterglow shock on
the outside and the expanding SN ejecta and central engine on the
inside.

\section{Pressure Confinement}\label{Widening}
Two assumptions made in sections \ref{Inside Star} and
\ref{EnvelopeConstraints} remain to be checked, both concerning the
effects of an external pressure on the cocoon or on the jet.  \S
\ref{S:PressureConf} addresses the assumption that the cocoon drives a
strong shock into the stellar envelope.  \S \ref{S:selfconf} concerns
the possibility that the jet is self-confined by its own cocoon
pressure, then spreads sideways outside the star before producing
gamma rays.

\subsection{Ambient pressure confinement} \label{S:PressureConf} 

If the ambient pressure is greater than the pressure within the
cocoon, then the assumption of a strong shock in equation
(\ref{betacPrelim}) is incorrect and the cocoon does not inflate as
described in \S \ref{cocoon}.
The cocoon is overpressured relative to the envelope by $(\beta_c
c)^2/c_s^2$, where $c_s$ is the envelope's isothermal sound speed: 
\begin{equation}\label{defAlpha}
c_s(r)^2 = \frac{p_a(r)}{\rho_a(r)}\equiv \alpha(r) \frac{GM(r)}{3r}. 
\end{equation}
With $\alpha(r)$ so defined, the virial theorem stipulates
$\left<\alpha\right>=1$ when the mean is weighted by binding energy
and $\left<1/\alpha\right>=1$ when weighted by thermal energy within
the star.  In general, $\alpha(r) \sim 1$ wherever the scale height is
of order the radius.  (In a polytrope of index $n$, $\alpha$ is
related to \citealt{C39}'s variable $v$ by $3/\alpha=(n+1)v$.)  Using
equations (\ref{betacSoln}) and (\ref{defAlpha}), $p_c>p$ when
\begin{eqnarray}\label{Pc>P}
L &\gtrsim& 
\frac{3\theta_j^{2/3}}{4\alpha} \left(\frac{c_s}{c}\right)^{14/3}
\frac{c^5}{G} 
\nonumber \\ &~& = 
\frac{\theta_j^{2/3} \alpha^{4/3} }{17.3} \left[\frac{G
M(r)}{r c^2}\right]^{7/3} \frac{c^5}{G}. 
\end{eqnarray}
This constraint is expressed in terms of the total jet luminosity $L$; 
I shall evaluate it at the fiducial collapse temperature $T_9\simeq 3.2$
\citep[O ignition;][]{1984ApJ...280..825B}. The appropriate value
of $\theta_j$, though different from $\theta_\gamma$, cannot exceed
unity. 

In ordinary core-collapse supernovae (below the pair instability
limit) the core is degenerate but the envelope above the collapsing
core is not; also, gas pressure dominates over radiation pressure.  At
the oxygen ignition temperature, condition (\ref{Pc>P}) becomes
\begin{equation}\label{Pc>P::Ordinary}
L \gtrsim 3.2  \times 10^{50} 
\alpha^{-1}
\theta_j^{2/3}
\left[\left(\frac{T_9}{3.2}\right)
\left(\frac{2}{\mu}\right)\right]^{7/3} ~\erg~\s^{-1}
\end{equation}
where $\mu$ is the mean mass per particle in a.m.u. 

In contrast to ordinary supernova cores, VMO cores are dominated by
radiation and are approximately $n=3$ polytropes in structure
\citep{1984ApJ...280..825B}.  They obey $R_{\rm core} = 0.26 [M_{\rm
core}/(100~\Msun)]^{1/2} (3.2/T_9) ~\Rsun$; hence
$p_c>p_a$ when
\begin{eqnarray}\label{Pc>P::VMO}
L &\gtrsim& 1.3 \times 10^{51}
 \alpha^{4/3}\theta_j^{2/3} \left(\frac{T_9}{3.2}\right)^{7/3}
 \nonumber \\ &~&\times 
\left(\frac{M_{\rm core}}{100~\Msun}\right)^{7/6}
 ~\erg~\s^{-1}.
\end{eqnarray}

The critical luminosities in equations (\ref{Pc>P::Ordinary}) and
(\ref{Pc>P::VMO}) should be compared to the value $L \simeq 3\times
10^{50} (10\%/\epsilon_\gamma) (10~\s/t_\gamma)~\erg~\s^{-1}$ implied
by \cite{2001ApJ...562L..55F}'s standard value $E_\gamma = 10^{50.5}$
erg.  In both ordinary supernovae and collapsing VMOs, the jet cocoon
could possibly be pressure-confined in the collapsing region ($T_9\gtrsim
3$), but not in the hydrostatic region where $T_9\lesssim 1$.

\subsection{Self-confinement by cocoon pressure}\label{S:selfconf}

The constraints on stellar envelopes derived above assume that
observational determinations of the intensity ($\Liso$) and opening
angle of gamma-ray emission can be applied to the earlier phase in
which the GRB crossed the stellar envelope.  This requires that the
jet is dynamically cold in the stellar interior, so that it does not
widen once outside the star.  It is important to consider the
alternative: that the jet may contain internal energy sufficient to
spread in angle after breakout. In this case the emission angle
$\theta_\gamma$ will exceed the jet angles $\theta_j$ attained within
the stellar envelope.  The jet would then be more intense within the
star than outside, vitiating (or at least relaxing) the constraints on
stellar envelopes derived in \S\S \ref{Inside Star} and
\ref{EnvelopeConstraints}.

For the flow to expand after breakout it must possess relativistic
internal energy, $h_j\rho_j \simeq 4p_j$, and it must be slow enough
to spread, $\Gamma_j<1/\theta_j$. (Note that cold jets will be heated
and decelerate locally to $\Gamma_j<1/\theta_j$, if they are crossed by an
stationary oblique shock.)
In this case the final opening angle will be set by the relativistic
beaming of the jet at breakout: $\theta_\gamma\simeq 1/\Gamma_j$;
a suggestion made by R. Blandford.  In general,
\begin{eqnarray}\label{ThetaPossibilities}
\theta_\gamma &\simeq &
\frac{1}{\Gamma_j(r_{\rm conf})}  + \theta_j(r_{\rm conf}) \nonumber
\\ &\simeq& \max\left[1/\Gamma_j(r_{\rm conf}),\theta_j(r_{\rm conf}) \right] 
\end{eqnarray}
where the two possibilities refer to hot, confined jets (with the
confining pressure is released at $r_{\rm conf}$) and cold, ballistic
jets, respectively.
\footnote{\cite{2003ApJ...586..356Z} suggest
eq. (\ref{ThetaPossibilities}) does not describe their numerical
results; however, it is not clear what $r_{\rm conf}$ or
$\Gamma_j(r_{\rm conf})$ is applicable to the wide-angle ejecta they
see. Although the choice of a single $r_{\rm conf}$ is perhaps
oversimplified, it remains a useful parameterization.}
One must have $r_{\rm conf}\simeq R$ for
pressure confinement to be significant; if $r_{\rm conf}\ll R$, the
ballistic-jet constraints of the previous section would still hold for
the outer envelope.  For confined jets, $\Gamma_j(R)$ is determined by
an observational determination of the final opening angle.  At the
same time, when $h_j \rho_j \simeq 4 p_j$, and so long as $\Gamma_j$
is still significantly above unity, the jet luminosity is
\[ L  = 4\pi R_j^2 \Gamma_j^2 p_j c ~~~~~~~~~~(\rm hot~jet) \] 
 (see the discussion after equation [\ref{LtildeDef}]), where
$R_j\equiv \theta_j r$ is the jet radius.  If evaluated at $r_{\rm
conf}$, $\Gamma_j$ may be replaced with $1/\theta_\gamma$.  

What determines $p_j$?  The confining pressure is clearly
time-dependent in this scenario, which complicates matters, but one
can set $p_j \simeq p_c$ at $r_{\rm conf}$.  The cocoon pressure is
$\rhoa (\beta_c c)^2$ where $\beta_c$ is given by equation
(\ref{betacSoln}):
\[ L^3 \rho_a = \pi^3 c^9 R_j^2 r^4 p_c^4, \] 
the primary assumption being that the jet head expands
nonrelativistically ($\Lt <1$).  Setting $p_j = p_c$ and using
$\Gamma_j = 1/\theta_\gamma$, 
\begin{equation}\label{Confinement}
2^4 (c R_j^2)^3 \pi \rho_a = \theta_\gamma^8 r^4 L ~~~~~ (\rm
  cocoon-confined,~ hot~jet)
\end{equation} 
which provides a means to evaluate pressure confinement if one has a
constraint on $R_j$.

\subsubsection{Adiabatic jets}\label{adiabatics}
One such constraint arises if the jet expands adiabatically from its
launching region, since this provides a formula for $R_j(p_j)$.  For
the steady flow of a relativistically hot gas ($p_j\gg \rho_j c^2$) 
at relativistic speeds ($\beta_j \simeq 1$) through a channel, the
simultaneous conservation of luminosity $L = (4 \Gamma_j^2 p_j) \pi
R_j^2 c$ and mass flux $\dot{M} = \Gamma_j \rho_j \pi R_j^2 c$, along with
the relation $p_j\propto \rho_j^{4/3}$, imply
\begin{equation}\label{AdiabaticJet}
\Gamma_j(r) = \frac{R_j}{r_0} ~~~~{\rm and}~~~~ p_j(r) =
\frac{L}{4\pi r_0^2 c} \Gamma_j^{-4} ~~~~~~ {\rm
  (adiabatic~jet)}
\end{equation}
where $r_0$ is the radius from which the flow accelerates through
$\Gamma_j=1$ and should reflect the dimensions of the central engine,
tens or hundreds of kilometers.  In equation (\ref{Confinement}), this
implies 
\begin{eqnarray}\label{AdConfinement}
\theta_\gamma &=& 1.6 \left(\frac{c^3 r_0^6 \rhoa }{L
  r^4}\right)^{1/14}
\nonumber \\ 
&=& 0.93^\circ \left(\frac{r_0}{10\,\rm km}\right)^{3/7} 
\left(\frac{\rhoa}{1\,\rm g\,cm^{-3} }\right)^{1/14} \nonumber \\ &~&~~~\times
\left(\frac{10^{52}\rm erg\,s^{-1}}{L}\right)^{1/14} 
\left(\frac{\Rsun}{r}\right)^{2/7}. 
\end{eqnarray} 
The $\gamma$-ray opening angle produced by pressure confinement of an
adiabatic jet is thus lower than any of the derived opening
angles, the discrepancy being much greater in terms of pressure (since
$p_j/p_c \propto \theta_\gamma^{-3.5}$ at fixed $r_0$). 

A jet expanding adiabatically from $r_0\sim$ 10 km can therefore be
confined in the core of the star, but would become cold and ballistic
in the outer envelope.  We found in \S \ref{Widening} that the ambient
core pressure may be high enough crush the cocoon, suggesting that the
jet can also be confined by ambient pressure in this region.  This
scenario was described by \cite{2001ApJ...556L..37M}; as they note, it
holds only in the stellar core and fails (producing a ballistic jet of
fixed $\theta_j = \theta_\gamma$) in the envelope.  However, we must
question whether $r_0$ might also change.

\subsubsection{Nonadiabatic jets: crossing shocks and mixing}\label{SS:mixing}
The above argument shows that a jet expanding from a launching region
$r_0$ of order 10 km will no longer be pressure-confined in the outer
stellar envelope.  But, what if the propagation is not adiabatic?
This can result from shocks that cross the jet, or from mixing between
the jet and its environment (cocoon or stellar envelope).
Heuristically one expects the jet's memory of $r_0$ to be erased in
either process; a larger effective $r_0$ would ameliorate the
discrepancy highlighted in \S \ref{S:selfconf}, perhaps making
pressure confinement more realistic.

First, consider a stationary oblique shock that crosses the jet.  The
postshock jet pressure ($p_{j2}$) is brought into equilibrium with
$p_c$, the external pressure that launches the shock.  The shock front
must be causal with respect to the postshock flow, i.e., sound travels
across the (postshock) jet faster than the shock itself does.  If the
postshock Lorentz factor is $\Gamma_{j2}$, sound fronts can propagate
at an angle $1/(\sqrt{3}\Gamma_j)$ to the jet axis. This must exceed
the angle $\theta_{cs}$ of the crossing shock relative to the jet axis
\citep[although usually not by much:][]{Whitham74}.  In order for
the shock to cross the jet, one must have $\theta_{cs}>\theta_j$; so,
$\Gamma_{j2}\lesssim 1/(\sqrt{3}\theta_{cs}) < \theta^{-1}$.  The jet
therefore makes a transition to a hot, pressure-confined state at its
current radius; the new effective value of $r_0$ for the postshock jet is
$R_j/\Gamma_{j2} \gtrsim \sqrt{3} \theta_{cs} R_j$. 

One can, in fact, estimate $\Gamma_{j2}$ and the shock obliquity
$\theta_{cs}$ from the expression $L_j = 4 \pi R_j^2 \Gamma_{j2}^2
p_{j2}c$ (valid for hot jets), from the condition $\Gamma_{j2}\lesssim
1/(\sqrt{3}\theta_{cs})$, and from the requirement that $p_{j2}\simeq
p_c$.  If $p_c = \beta_c^2 \rhoa c^2$ is estimated via equation
(\ref{betacSoln}), one finds
\begin{equation}\label{eq:thetaCS}
\frac{\theta_{cs}}{\theta_j} \simeq 3.5
\left[\frac{\Menv}{\Msun}\frac{\Rsun}{R}
\frac{10^{52}\rm erg/s}{L_j}
\left(\frac{3^\circ}{\theta_j}\right)^2\right]^{1/8}, 
\end{equation}
which can be read as an estimate of the number of crossing shocks
along the jet axis.  Note that eq. (\ref{ThetaPossibilities}) predicts
$\theta_\gamma \simeq \theta_{cs}$, roughly speaking, if the jet is
effectively released after one of its crossing shocks. These results
appear broadly compatible with the work of
\cite{2003ApJ...586..356Z}, and indicate the potential importance
of this route to pressure confinement.

Second, consider the mixing of ambient material into the jet during a
pressure-confined phase.  Nonadiabatic mixing can be described by
imagining relativistically hot jet material flowing though a uniform
channel of fixed cross-sectional area $A_j$. At some initial time the
channel is allowed to mix with some area $d A_j$ of ambient material
(density $\rho_c$, pressure $p_c$) at rest. Assuming perfect mixing
and nonrelativistic surroundings ($p_c\ll \rho_c c^2$), the change in
jet parameters $p_j$, $\rho_j$, and $\Gamma_j$ are determined by the
conservation of energy, momentum, and rest mass through the mixing
event. That is,
\begin{equation}\label{mixing1}
 d \left(  \left\{\begin{array}{cc} 
\Gamma_j^2 h_j\rho_j -p_j \\                    
\Gamma_j^2 \beta_j h_j\rho_j \\                 
\Gamma_j \rho_j                           
\end{array} \right\} A_j\right) =
\left\{\begin{array}{cc} 
\rho_c c^2 \\                              
0          \\                              
\rho_c                                     
\end{array} \right\} d A_j. 
\end{equation}
Algebra then yields
differential equations for $p_j$, $\rho_j$, and $\Gamma_j$ as
functions of $A$. 

A couple interesting conclusions can be drawn from this exercise.
First, the heuristic argument that $r_0$ would be forgotten is
correct: the effective value of $r_0 \equiv R_j/\Gamma_j$ increases
monotonically in non-adiabatic jet expansion, because $R_j$ increases
whereas $\Gamma_j$ decreases (or if $\rho_c =p_c=0$, stays constant).
This in turn implies that pressure confinement is extended somewhat by
mixing.

\begin{figure*}
\centerline{\epsfig{figure=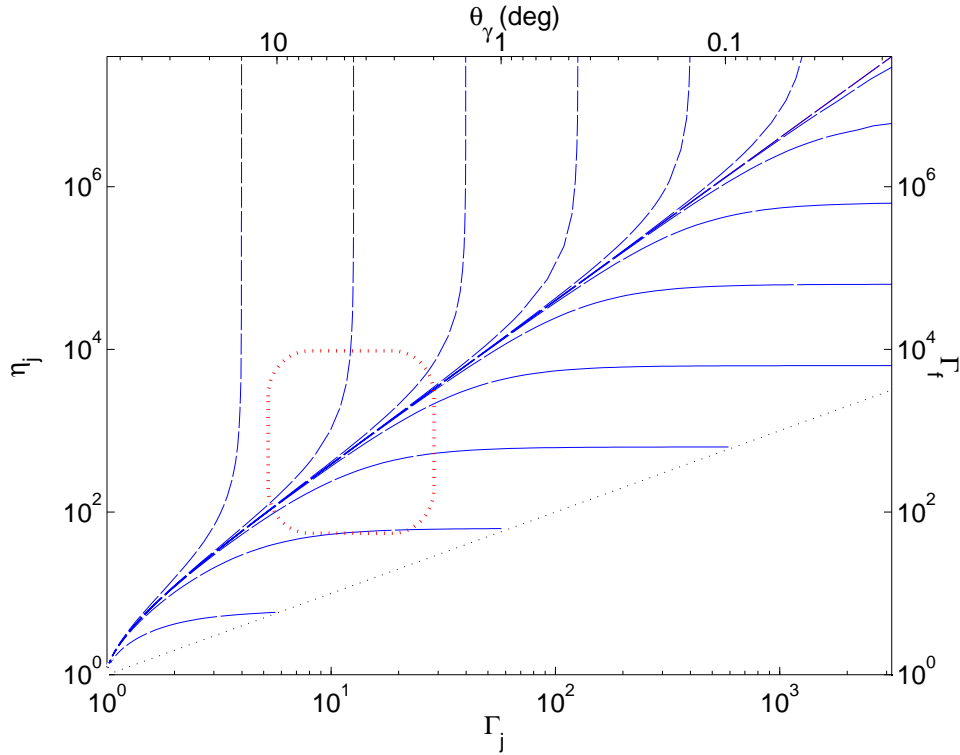,height=4in}} 
\caption[Evolution of jets under mixing] {\footnotesize Evolution of
pressure-confined jets due to mixing with their surroundings
(eqs. [\ref{MixingGamma}] and [\ref{MixingEta}]).  The ratio $\eta_j$
of luminosity to mass flux is plotted versus jet Lorentz factor; these
set the final Lorentz factor $\Gamma_f$ and opening angle
$\theta_\gamma$.  Breaks in the trajectories are spaced by a doubling
of jet mass per unit length.  Mixing causes jets to move down and to
the left along the plotted trajectories toward $\eta_j=\Gamma_j =1$.
In contrast, adiabatic evolution causes jets to move horizontally
rightward to higher $\Gamma_j$ (smaller $\theta_\gamma$) at fixed $\eta_j$
and $\Gamma_f$.  No solutions exist below the dotted line
($\eta_j=\Gamma_j$), where pressure is negative.  Accepted ranges of
$\theta_\gamma$ and $\Gamma_f$ are estimated by the dotted rectangle.
Inside this region, mixing correlates observables: higher $\Gamma_f$
corresponds to lower $\theta_\gamma$.
\label{figMix} }
\end{figure*}

Secondly, mixing forces the nondimensional jet parameters to trace a
characteristic trajectory that whose implications would be observable.
Recall that the final opening angle $\theta_\gamma$ is the inverse of
$\Gamma_j$ if the jet is confined.  Also, final Lorentz factor
$\Gamma_f$ is equal to the ratio of the jet's luminosity to its mass
flux, \[ \eta_j \equiv \frac{L_j}{\dot{M}_j c^2} =
\left(\frac{4p_j}{\rho_j c^2}+ 1\right)\Gamma_j.\]
In terms of these dimensionless parameters, equation (\ref{mixing1})
can be restated 
\begin{eqnarray}\label{MixingGamma}
\frac{d\Gamma_j}{d\ln M_j} &=&
-\Gamma_j^2\frac{(4\Gamma_j^2-1)(\Gamma_j^2-1)}
{\eta_j(2\Gamma_j^2+1) + \Gamma_j^3-\Gamma_j};
\\  \label{MixingEta}
\frac{d\eta_j}{d\ln M_j} &=& 
-\eta_j \frac{\eta_j + 2\Gamma_j^2(\eta_j-2) + \Gamma_j^3}
{\eta_j(2\Gamma_j^2+1) + \Gamma_j^3-\Gamma_j},  
\end{eqnarray} 
where $d\ln M_j = \rho_c dA/(\Gamma_j \rho_j A)$ represents the
fractional mass per unit length mixed in.  The ratio of these
equations (in which the denominators cancel) describes set mixing
trajectories for $\eta_j(\Gamma_j)$, as shown in fig. \ref{figMix}.
There exists an attractor solution, which is $\eta = 4\Gamma_j^2$
(i.e., $p_j = \Gamma_j \rho_j c^2$) when $\Gamma_j\gg 1$ and
approaches $(\eta_j,\Gamma_j)\rightarrow (1,1)$. Mixing trajectories
rapidly join onto it by decreasing either $\Gamma_j$ or $\eta_j$ as
mass is mixed in.

As shown in figure \ref{figMix}, mixing trajectories do cross through
reasonable values of $\Gamma_f$ and $\theta_\gamma$. Does this mean
that these jet parameters arise from mixing?  This seems plausible if
there ever exists phase of pressure-confined evolution in the stellar
core or mantle.  If $2\Gamma_j<\eta_j^{1/2}$, adiabatic expansion and
mixing both act to bring $2\Gamma_j\rightarrow \eta_j^{1/2}$. If
$2\Gamma_j> \eta_j^{1/2}$, the adiabatic tendency for $\Gamma_j$ to
increase is counterbalanced by the nonadiabatic decrease in
$\Gamma_j$.  This makes the mixing attractor solution a probable one
for the end state of jets that are pressure-confined at some stage.  (A
more detailed analysis would require a dynamical analysis of jet expansion.)

If so, then the attractor solution would imprint the correlation
\begin{equation}\label{MixingAttractor}
\theta_\gamma\simeq 11^\circ \left(\frac{100}{\Gamma_f}\right)^{1/2} ~~~~
   {\rm (mixing~attractor)}
\end{equation} 
which may have observational implications in that $\Lisogam\propto
\Gamma_f$ for a fixed $L_j/\epsilon_\gamma$.  Such a trend has been
implicated in the correlations between spectral lag and luminosity,
variability and luminosity, and afterglow break time and luminosity
among GRBs \citep[e.g.,][]{2002NewA....7..197R,2002ApJ...569..682S}.  
The mixing attractor provides a motivation for a rather tight
correlation among bursts with similar jet luminosities. 

\subsubsection{Confined jets: timing properties}
The variability-luminosity correlation would be affected also by the
filtering of jet variations by pressure gradients, which occurs only
when the jet is pressure-confined.  Now, the interal dynamical time of
a relativistic jet is $\sim R_j/c$, or $\Gamma_j R_j/c$ in the lab
frame.  If a strobe approached the observer at $\Gamma_j$, pulsating
once per dynamical time, the observed period would be foreshortened to
\begin{equation} \label{deltaTobsPressure}
\delta \tobs \simeq \frac{R_j}{\Gamma_j c} = \frac{r_0}{c}. 
\end{equation}  
The same result can be derived by considering a standing-wave pattern
of wavelength $R_j$ in the jet's frame, or $R_j/\Gamma_j$ in the lab
frame, that is swept past a fixed pressure-release radius $r_{\rm
conf}$ and leads to jet pulsations as the peaks and troughs pass by.
As we saw above, crossing shocks and mixing increase the effective
value of $r_0$ and thus increase the characteristic variability
timescale.  Shocks and mixing also decrease $\Lisogam$ by decreasing
$\Gamma_j$ and thereby increasing $\theta_\gamma$; this implies a correlation between jet variability and
$\gamma$-ray brightness.

\subsection{Pressure Confinement and Envelope Constraints}
\label{confinementevaluation} 
What should we make of the possibility that pressure-confined jets
will evade the constraints on stellar envelopes derived in \S\S
\ref{Inside Star} and \ref{EnvelopeConstraints} for ballistic jets?
Equation (\ref{AdConfinement}) shows that jet confinement works if
\[ r_0  = 500 \left(\frac{\theta_\gamma}{5^\circ}\right)^{7/3} 
\left(\frac{r_{\rm conf}}{\Rsun}\right)^{2/3} \left(\frac{1 \,\rm
g\,cm^{-3}}{\rho_a}\right)^{1/6} L_{52}^{1/6} {\, \rm km}, \] a value
that does not appear to violate timing constraints
(eq. [\ref{deltaTobsPressure}]).  This indicates pressure-confined
jets are plausible if shocks or jet mixing increase $r_0$ sufficiently. 

Why would a jet change from pressure-confined to ballistic?  This
could occur in a couple ways.  First, the jet could accelerate in such
a way that sound no longer crosses it despite it being hot, i.e.,
$\Gamma_j(r) > 1/\theta_j(r)$ for some range of $r$.  However this
appears to require the ambient density to drop to a near vacuum,
implying it would not occur spontaneously within a star.
Alternatively, the jet could become internally nonrelativistic
($p_j\lesssim \rho_j c^2$). As we have seen, adiabatic evolution works
in this direction whereas crossing shocks and non-adiabatic mixing
oppose it.  The transition would then depend on the specifics of jet
propagation and mixing.  Finally, the jet could be confined by
internal magnetic stresses, a possibility highlighted by the
$\gamma$-ray polarization discovered by \cite{2003Natur.423..415C},
but beyond the scope of this paper.

Jet pressure confinement would naturally lead to time-dependent jet
properties, since the confining pressure evolves during the cocoon
phase and after breakout.  Crossing shocks and mixing with ambient
material are also unlikely to be steady processes.  One would expect
evolution during GRBs as a result; however, no trends are discernible
except for changes in pulse asymmetry \citep[][]{2000ApJS..131....1L}.

Apart from this contraindication, it is difficult to judge whether
pressure confinement persists across entire stellar radii.  If so,
then the constraints on stellar envelopes in \S\S \ref{Inside Star}
and \ref{EnvelopeConstraints} are relaxed.  Numerical simulations can
potentially solve detailed questions such as this one.  However,
caution should be used in interpreting them, because of the many
decades separating the launching radius from the stellar radius.

\section{Conclusions}\label{Conclusions}
This paper constrains possible stellar progenitors for GRBs by
requiring that jets of GRB-like luminosities and durations can clear a
path for themselves in the star's envelope prior to producing gamma
rays.  Of all the possible progenitors, only the compact carbon-oxygen
post-Wolf-Rayet stars (SN Ic progenitors) and the bare cores of very
massive objects can plausibly collapse in the durations of long GRBs
at low redshift (\S \ref{S:freefalltimes}, figures \ref{fig1} and
\ref{fig2}).  In more extended stars, the outer stellar envelope
remains to impede the progress of the jet, and it is likely to violate
observational constraint.  Therefore, I conclude that GRBs with SN
progentiors come primarily from type Ic or VMO-core events.  Type Ib is
not thoroughly ruled out by this work, but type II (supergiant stars) are.

The most stringent constraint on a stellar envelope arises from the
requirement that the jet can traverse it in an observed time not much
longer than the duration of the GRB.  Under the assumption of a
ballistic rather than a pressure-confined jet, this constraint is
independent of the inferred opening angle of the burst, and (given an
observed fluence and duration) depends on its inferred comoving
distance rather than its luminosity distance.  This makes it
insensitive to an uncertainty in redshift.  Given the luminosities and
durations of GRBs (regardless of their redshift;
fig. \ref{figTiming}), only post-Wolf-Rayet stars and VMO cores are
compact enough to satisfy this criterion.  The variability-luminosity
correlation discussed by \cite{GRBcepheid} and
\cite{2001ApJ...552...57R} allows this constraint to be applied to a
large number of bursts in the BATSE catalog (figure \ref{figTiming}).
Only a few are compatible with blue supergiant progenitors; red
supergiants and VMOs with envelopes are ruled out.  Also ruled out is
the ``supranova'' model of \cite{1999ApJ...527L..43V} (a supernova
followed by a GRB), unless the SN ejecta are optically thin by the
time of the GRB, or unless the pulsar nebula is energetic enough to
clear them aside.  

Post-Wolf-Rayet stars have been favored among stellar GRB progenitors
since the work of \cite{1999ApJ...524..262M}, on the basis that their
compact envelopes delay the GRB jet breakout the least.  The above
constraint quantifies and strengthens this conclusion, and relates it
to the observed properties of GRBs rather than those of a specific
model for the central engine.

For ballistic jets, an additional constraint arises from the
requirement that the GRB jet cocoon should not overtake its driving
jet and produce a spherical explosion.  This is generally not as
restrictive as the constraint from burst durations discussed above,
but it does become important for jets with opening angles exceeding
$10^\circ$ (figure \ref{figCocoon}).  VMO cores, being the most
compact, are the progenitors most likely to be ruled out this way if
they have not collapsed prior to the GRB.  The requirement of a
jet-cocoon structure also gives interesting upper and lower bounds on
the energy entrained in the stellar envelope during the phase of jet
propagation (eq. [\ref{EinConstraint}]). This energy is stored in the
jet cocoon and is available to drive a ``dirty'' fireball
\citep{1998ApJ...494L..45P} of expanding cocoon material after the jet
breaks out.

The breakout phenomenon is itself a candidate for producing a
transient, as may have happened in SN 1998bw to produce GRB 980425.
In section \ref{Breakout} it is calculated how much kinetic energy is
channeled into relativistic envelope ejecta during a jet's breakout,
by matching the propagation law for the jet's terminal shock onto the
relativistic shock and post-shock acceleration behavior described by
\cite{2001ApJ...551..946T}.  I find the energy of this ejecta to be
small compared to that of the burst.  \cite{2002MNRAS.331..197R} have
recently discussed how the upscattering of photons from the shocked
envelope by the jet may produce a precursor of hard gamma rays;
however, note that this is reduced in importance for the compact
stellar progenitors favored by the timing constraints.
\cite{2003ApJ...586..356Z} have suggested breakout ejecta as the
origin of short, hard bursts; however their estimate of the energy is
at odds with that calculated here.

If the gamma-ray photons are not able to clear away a stellar wind in
the region around the star in the manner described by
\cite{2000ApJ...534..239M}, \cite{2000ApJ...538..105T}, and
\cite{beloborodov-sweeping}, then the presence of this wind places an
upper limit on the jet Lorentz factor.  This limit arises in the
internal shock model for GRB emission because the presence of the
external shock limits the distance within which internal shocks can
form.  The equivalent limit has been presented previously for uniform
ambient media \cite{1999PhR...314..575P}; however, for a sufficiently
dense stellar wind it can conflict with the lower limits on jet
Lorentz factor \citep[e.g.,][]{2001ApJ...555..540L}. 

\begin{figure*}
\centerline{\epsfig{figure=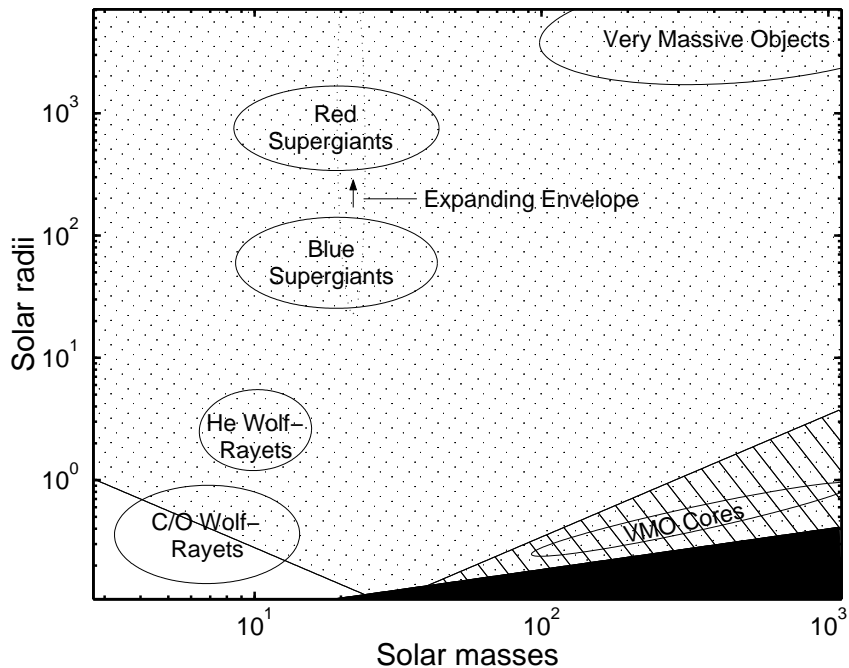,height=3.5in}} 
\caption[Constraints from GRB 000418] {\footnotesize Constraints on
possible stellar progenitors for the case of GRB 000418 (assuming a
ballistic jet and $\varepsilon_\gamma = 10\%$).  Stars in the {\em
black shaded region} would collapse in the seven-second intrinsic
duration of the burst. Those within the {\em hatched region} are
excluded because they are too dense for a jet-cocoon structure to
exist given the luminosity of this burst, and would develop a
spherical blastwave instead.  Those in the {\em dotted region} are
also excluded because the GRB jet would take much longer than the
observed duration to traverse their envelopes.
\label{fig418} }
\end{figure*}

Figure \ref{fig418} illustrates the above criteria for the specific
case of GRB 000418, assuming a ballistic jet with $\varepsilon_\gamma
= 10\%$.  Its seven-second duration is briefer even than the free-fall
times of VMO cores.  Because of its large opening angle
\citep[$11^\circ$;][]{2001ApJ...556..556B}, it would not succesfully
form a jet-cocoon structure in any uncollapsed portion of the VMO
core.  In fact, a jet of its inferred luminosity could cross nothing
more extended than the most compact of Wolf-Rayet stars in the GRB
duration. I conclude from this that it came either from a compact
carbon-oxygen Wolf-Rayet star, or from a VMO core that managed to
produce a GRB of briefer duration than its free-fall time, or that it
did not have a supernova origin. These restrictions change
quantitatively, but not qualitatively, if $\varepsilon_\gamma \ll
10\%$.

Pressure confinement of jets provides a way for GRB hosts to evade the
constraints on stellar envelopes listed above for ballistic jets.
This occurs because, for a given $\gamma$-ray opening angle, confined
jets are narrower and more intense than their ballistic counterparts.
The computational results of \cite{2000ApJ...531L.119A} and
\cite{2003ApJ...586..356Z} correspond to pressure-confined jets.  In \S
\ref{adiabatics} it was shown that jets that do not mix with their
environs cannot remain pressure-confined \citep[see
also][]{2001ApJ...556L..37M}.  Simulations that do not resolve the
jets' launching scale may not observe this effect, however.  Mixing
of jet and envelope (\S \ref{SS:mixing}) is capable of sustaining a
pressure-confined state. 

Intriguingly, mixing of jet and envelope imprints on the jet a tight
correlation (the ``mixing attractor'' of eq. [\ref{MixingAttractor}]
and figure \ref{figMix}) between its Lorentz factor and energy per
unit mass.  This leads to a correlation between final opening angle and
Lorentz factor, which in turn may be related to the lag-luminosity and
variability-luminosity correlations observed in burst catalogs.
Excessive mixing, however, has the effect of filtering rapid
fluctuations from jets. 

This work was stimulated by a visit to UC Santa Cruz and by
interactions with Stan Woosley, Andrew MacFadyen, Alex Heger, and
Weiqun Zhang while I was there. It was further motivated by
conversations with and encouragement from Roger Blandford, Re'em Sari,
Sterl Phinney and Josh Bloom during a visit to Caltech. It is a
pleasure to thank Woosley, Blandford, and Phinney for their
hospitality during those visits.  I am especially grateful to Nicole
Lloyd-Ronning for explaining the GRB luminosity-variability relation
and suggesting clarifications.  I also thank Andrei Beloborodov and
Chris Thompson for discussing jet-envelope interactions and Chris
Fryer for discussing progenitor scenarios.  Comments from Beloborodov,
Charles Dermer, Jonathan Tan, Chris McKee, John Monnier, and the
referee, Ralph Wijers, are also appreciated.  This work was supported
by NSERC and by the Canada Research Chairs program.


\label{lastpage}

\end{document}